\documentclass[reprint,superscriptaddress,amsmath,amssymb,amsthm,aps,prx,nofootinbib]{revtex4-2}

\usepackage{graphicx}
\usepackage{bm}
\usepackage{enumitem}
\usepackage[version=4]{mhchem}
\usepackage{inconsolata}
\usepackage[ruled,vlined,longend,algo2e]{algorithm2e}
\usepackage{titletoc}
\usepackage{setspace}

\usepackage[table,dvipsnames]{xcolor}
\definecolor{linkcolor}{HTML}{b91c1c}
\definecolor{citecolor}{HTML}{52525b}
\definecolor{urlcolor}{HTML}{1d4ed8}

\usepackage{booktabs}
\usepackage{tabularray}
\UseTblrLibrary{booktabs}
\usepackage{makecell}
\usepackage{multirow}

\usepackage{hyperref}
\hypersetup{colorlinks,linkcolor=linkcolor,citecolor=citecolor,urlcolor=urlcolor}
\bibpunct{\color{citecolor}[}{\color{citecolor}]}{,}{n}{}{;}  
\usepackage[capitalise]{cleveref}

\newcommand{\siref}[1]{Supplementary Note~#1}

\usepackage{amsthm}

\crefname{extendedfigure}{Extended Data Fig.}{Extended Data Figs.}
\crefname{extendedtable}{Extended Data Table}{Extended Data Tables}

\begin{document}

\title{Universal Thermodynamic Interatomic Potentials for Crystalline Materials}

\author{Juno Nam}
\affiliation{Department of Materials Science and Engineering, Massachusetts Institute of Technology, Cambridge, MA 02139, USA}
\author{Bowen Deng}
\affiliation{Department of Materials Science and Engineering, Massachusetts Institute of Technology, Cambridge, MA 02139, USA}
\author{Xiaochen Du}
\affiliation{Department of Materials Science and Engineering, Massachusetts Institute of Technology, Cambridge, MA 02139, USA}
\author{Luis Barroso-Luque}
\affiliation{Fundamental AI Research, Meta, San Francisco, CA 94105, USA}
\thanks{Meta-affiliated authors served solely in an advisory role. All access to the models, datasets, and code, as well as the development and release of the dataset and code, was carried out exclusively by the non-Meta authors or their academic institution.}
\author{Benjamin Kurt Miller}
\email{bkmi@meta.com}
\affiliation{Fundamental AI Research, Meta, San Francisco, CA 94105, USA}
\thanks{Meta-affiliated authors served solely in an advisory role. All access to the models, datasets, and code, as well as the development and release of the dataset and code, was carried out exclusively by the non-Meta authors or their academic institution.}
\author{Rafael G\'omez-Bombarelli}
\email{rafagb@mit.edu}
\affiliation{Department of Materials Science and Engineering, Massachusetts Institute of Technology, Cambridge, MA 02139, USA}

\date{\today}

\begin{abstract}
Free energies govern solid-state phase stability, yet computational materials discovery still relies largely on ground-state energies because free energy calculations require ensemble averages.
We introduce the \emph{thermodynamic interatomic potential} (TIP), which extends an interatomic potential from its static energy to a thermodynamically consistent Gibbs free energy model, with thermodynamic responses following from temperature and pressure by automatic differentiation.
We implement TIP[UMA] using the universal potential UMA, train it on free energies from quasi-harmonic to molecular dynamics fidelity, and calibrate it to higher-resolution calculations or experiment.
From a single evaluation, it returns the equation of state of a crystal and locates phase transitions among competing branches, including dynamically stabilized phases.
Fine-tuning extends the model to alloy solubility limits and miscibility gaps.
TIP makes the free energy as accessible as the potential energy, opening finite-temperature phase stability to high-throughput discovery.
\end{abstract}

\maketitle


\section{Introduction}
\label{sec:introduction}

Phase diagrams, stability fields, and metastability windows in solid-state materials science are determined by free energies at the relevant conditions, not by ground-state energies alone.
Nevertheless, computational materials discovery still relies heavily on static density functional theory (DFT) energetics and convex-hull constructions, because large-scale screening workflows were built around ground-state data~\citep{jain2013commentary, kirklin2015open, sun2019map}.
This creates a persistent imbalance: structure and energy predictions are available at large scale, whereas the temperature- and pressure-dependent free energies remain less available, more computationally expensive, and less standardized~\citep{tolborg2022free}.

This imbalance arises because a ground-state energy characterizes one relaxed configuration, whereas a free energy averages over a thermodynamic ensemble including entropic contributions.
Computing a free energy requires averaging over thermally accessible configurations for each structure and thermodynamic condition.
Approximating this ensemble can be inaccurate because vibrational and configurational contributions often determine phase stability, and vibrational free energies alone can shift phase boundaries by amounts comparable to configurational contributions~\citep{garbulsky1996contribution, van2002effect}.
The vibrational contribution is particularly demanding because it is strongly temperature dependent and deviates increasingly from harmonic behavior as anharmonicity grows.
Phase stability depends on small free-energy differences between competing phases and decomposition products, rather than only on formation from the elements~\citep{bartel2019role, bartel2020critical}.
Therefore, these free energies must be resolved precisely, over the temperature and pressure ranges in which the solid phases remain stable or metastable~\citep{sun2016thermodynamic, aykol2018thermodynamic}.

First-principles thermodynamics has been addressing this burden by coarse-graining free energies into tractable effective models.
The cluster expansion~\citep{sanchez1984generalized, connolly1983density, zhong2022l0l2} represents the configurational energy of a crystal on a fixed parent lattice as a sum of effective interactions fitted to first-principles energies, and statistical-mechanical calculations then yield composition--temperature phase behavior~\citep{asta1992first, de1994cluster}.
This approach applies when a crystal can be represented as occupations on a parent lattice~\citep{ceder2000first, van2002automating}, but its interactions are specific to a chemistry and lattice and must be refitted for a new system.
Additionally, the fitted quantity is usually configurational energy, and temperature enters through a separate statistical-mechanical calculation rather than as model inputs.

A separate hierarchy addresses the vibrational free energy of a fixed crystal.
Quasi-harmonic lattice dynamics, built on density functional perturbation theory~\citep{baroni2001phonons} and finite-displacement phonon workflows~\citep{togo2015first}, is the default route to solid-state free energies.
Because it absorbs anharmonicity only into the volume dependence of harmonic phonons, it grows fragile in the regimes that govern many functional solids: soft modes, strong phonon renormalization, entropy-stabilized high-temperature phases, and mechanically unstable harmonic references.
Self-consistent phonon theories~\citep{souvatzis2008entropy}, temperature dependent effective potential methods~\citep{hellman2013temperature}, and formalisms for unstable phases~\citep{thomas2013finite, van2015free} address these cases.
At higher fidelity, explicit free energy integration evaluates absolute free energies based on molecular dynamics (MD) simulations, where thermodynamic integration proceeds through Frenkel--Ladd switching~\citep{frenkel1984new}, reversible scaling~\citep{de1999optimized}, and alchemical switching~\citep{nam2025alchemical}, realized as nonequilibrium MD workflows~\citep{menon2021automated, cheng2018computing}.
The computational cost increases with the degree of anharmonicity resolved, and is especially high when energies and forces are evaluated from first principles.

Universal machine learning interatomic potentials (MLIPs)~\citep{deringer2019machine, behler2007generalized, bartok2010gaussian, chen2022universal, deng2023chgnet, batatia2025foundation, mazitov2025pet} reduce this computational cost by reproducing first-principles energies and forces across broad chemical spaces, enabling high-throughput relaxation and large-scale MD without system-specific refitting.
However, MLIPs predict the potential energy of individual configurations.
Free energy calculations still require ensemble sampling for each structure and condition, even though individual force evaluations are computationally inexpensive.

An alternative is to model free energy directly rather than apply a statistical-mechanical calculation to an energy model.
The CALPHAD method~\citep{kaufman1970computer, lukas2007computational} represents the Gibbs free energy of each phase as an analytic function of temperature and composition, assessed from experimental and first-principles data.
These assessments are semi-empirical, specific to each chemical system, and often proprietary.
Descriptor-based Gibbs free energy model~\citep{bartel2018physical} predicts temperature-dependent free energies of stoichiometric solids using compact, interpretable composition-level descriptors, but they are limited to near-ambient pressure and lack the structural resolution needed to identify phase transitions.
Structure-resolved surrogates have been developed for specific systems, such as a machine-learned metastable phase diagram of carbon~\citep{srinivasan2022carbon}, but remain confined to a single chemistry or a narrow class of structures.
A chemically transferable, branch-resolved Gibbs free-energy model is still needed, which spans harmonic and anharmonic fidelity levels and derives thermodynamic response functions from one differentiable surface.

\begin{figure*}[tp]
\centering
\includegraphics[width=\textwidth]{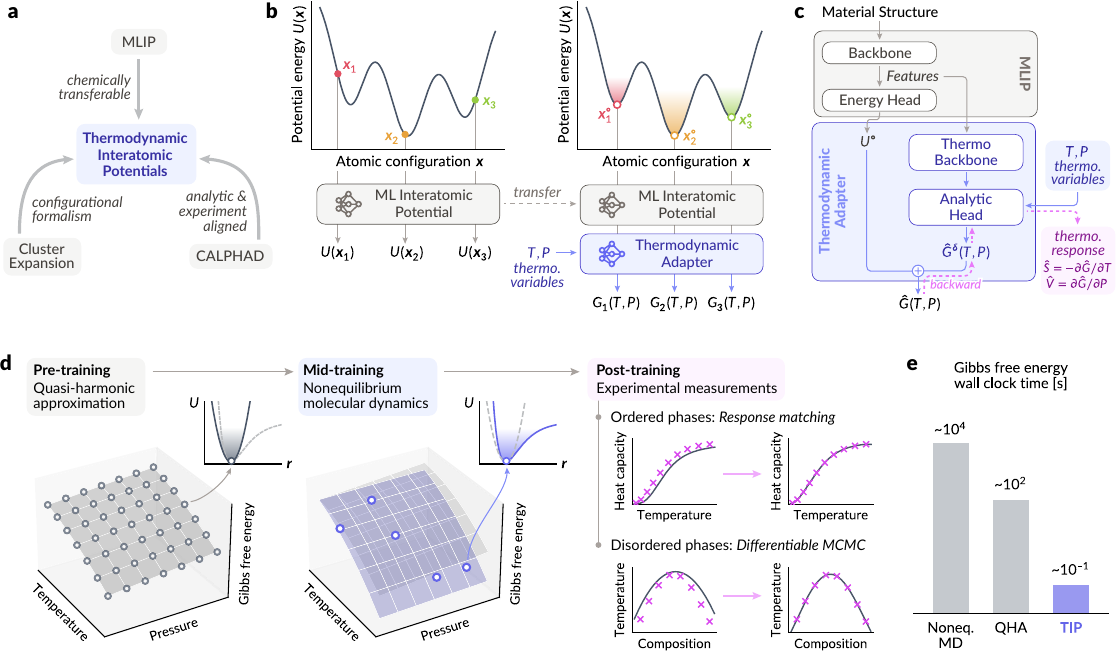}
\caption{%
\textbf{Free energy adaptation with a thermodynamic interatomic potential (TIP).}
\textbf{(a)} A TIP combines three established routes to solid-state thermodynamics: the chemical transferability of MLIPs, the configurational phase stability formalism of the cluster expansion, and the analytic, experiment-aligned free energy representation of CALPHAD.
\textbf{(b)} A TIP augments a pre-trained MLIP with a thermodynamic adapter: for each locally relaxed branch representative $\bm{x}^\circ$, the frozen MLIP encodes $\bm{x}^\circ$, and the adapter combines those features with $(T,P)$ to predict the branch Gibbs free energy $\hat{G}(T,P)$.
\textbf{(c)} Architecture: the MLIP backbone's energy head returns the static reference $U^\circ := U(\bm{x}^\circ)$ and its per-atom features condition the adapter, whose analytic head outputs the thermodynamic residual $\hat{G}^\delta(T,P)$. Their sum defines $\hat{G}$, and its derivatives give the response properties.
\textbf{(d)} Three-stage training: pre-training on quasi-harmonic $(T,P)$ grids, mid-training on anharmonic absolute free energies from nonequilibrium MD, and post-training against experimental data by response matching for ordered phases (e.g., heat capacity) and differentiable reweighting for disordered phases (e.g., composition--temperature binodals).
\textbf{(e)} Wall-clock time per Gibbs free energy evaluation on a given $\bm{x}^\circ$: nonequilibrium MD (${\sim}10^4$~s) and the quasi-harmonic approximation (${\sim}10^2$~s) compared with one TIP[UMA] evaluation (${\sim}10^{-1}$~s), five and three orders of magnitude faster.
}
\label{fig:overview}
\end{figure*}

We therefore introduce the thermodynamic interatomic potential (TIP).
Conceptually, TIP integrates three complementary ideas from existing approaches to solid-state thermodynamics (\cref{fig:overview}a): it inherits the chemical transferability of MLIPs; represents the thermodynamics of an ordered branch with an efficient surrogate, analogous to the cluster expansion but without lattice- or chemistry-specific parameterization; and adopts the CALPHAD philosophy of modeling free energy directly as a differentiable thermodynamic function.
Together, these elements enable equilibrium properties and thermodynamic response functions to be predicted without repeated statistical-mechanical calculations.

For a relaxed ordered bulk crystal, TIP encodes the structure once and returns a differentiable approximation to that branch's Gibbs free energy surface across a range of temperature and pressure in a single evaluation.
This confines the statistical-mechanical cost to training, so the free energy of a new crystal within the model's domain follows without further sampling.
TIP resolves the vibrational thermodynamics of one ordered branch at a time, and configurational thermodynamics is obtained by sampling ordered branches across composition, extending the model to solid solutions and miscibility gaps.
The current model identifies transitions only among the supplied branch candidates, i.e., it does not discover crystal structures.
Also, the liquid state is outside the present scope because it has no ordered branch representation.
TIP is built on a universal interatomic potential and trained sequentially on quasi-harmonic and MD free energies, and experimental calibration data.
We demonstrate prediction of equations of state and high-temperature phase transitions, including phases without stable harmonic references.

\section{Results}
\label{sec:results}

\subsection{Thermodynamic Interatomic Potentials}

TIP predicts the Gibbs free energy of an ordered crystal as a continuous function of temperature and pressure, from its relaxed structure (\cref{fig:overview}b).
While conventional free energy calculations require thermal ensemble sampling and thermodynamic path integration at every state point, TIP instead evaluates a fitted continuous surface.
We enable this via \emph{free energy adaptation}: the absolute Gibbs free energy of a crystal is decomposed into its static ($0$~K) energy plus a smooth \emph{thermodynamic residual}.
A pre-trained MLIP supplies the static relaxed energy $U^\circ$, so the adapter model learns only the residual $\hat{G}^\delta(\bm{x}^\circ;\,T,P)$, giving the predicted Gibbs free energy
\begin{equation}
\label{eq:G-split-main}
    \hat{G}(\bm{x}^\circ;\,T,P) = U^\circ + \hat{G}^\delta(\bm{x}^\circ;\,T,P).
\end{equation}
Learning $\hat{G}^\delta$ rather than $\hat{G}$ reduces the structure-dependent dynamic range of the target and focuses the adapter on finite-temperature contributions.
A single relaxed structure then generates an entire free energy surface over $(T,P)$.

The frozen MLIP encodes the structure, and the adapter uses its per-atom features and the thermodynamic variables $(T,P)$ as inputs (\cref{fig:overview}c).
The thermodynamic backbone is temperature- and pressure-agnostic: from the structure features alone, it predicts the coefficients of a closed-form thermodynamic expression, and $(T,P)$ enter only through the analytic head.
The functional forms are drawn from established physical models: the temperature dependence combines a quantum Einstein oscillator vibrational free energy with a low-order polynomial, as in CALPHAD Gibbs energy functions~\citep{kaufman1970computer, lukas2007computational}, and the pressure--volume relation of each per-atom contribution follows a Murnaghan-like equation of state~\citep{murnaghan1944compressibility}.
Because these forms are analytic, the fitted coefficients carry physical meaning, and derivatives of the free energy, including higher-order responses, remain smooth by construction.
Furthermore, automatic differentiation yields thermodynamic conjugate variables such as volume and entropy, and thermodynamic responses, such as heat capacity, bulk modulus, and thermal expansion, from the same $\hat{G}$, ensuring thermodynamic consistency within the model.
Because the analytic head assembles the free energy from per-atom contributions, model inference requires only the relaxed unit cell, even though the training labels are computed on the large supercells required to converge phonon and MD free energies.
TIP is therefore trained to reproduce those free energies from a unit cell input, at a computational cost set by the unit cell rather than the simulation supercell.
This reduces calculations requiring hours of explicit sampling or minutes of a quasi-harmonic workflow to seconds of model inference (\cref{fig:overview}e).

TIP models one ordered crystalline branch at a time, represented by its relaxed representative $\bm{x}^\circ$.
A branch may be a basin of the potential energy surface, a symmetry-constrained stationary point for a dynamically stabilized phase (\cref{fig:overview}b), a distinct polymorph, or a distinct chemical ordering.
Each branch has a distinct predicted free energy surface $\hat{G}(\bm{x}^\circ;\,T,P)$.
Stable phases and the transitions between phases are then resolved by comparing the predicted free energies of the candidate branches.
The formal branch construction and its underlying assumptions are given in \siref{A}.
This way, a TIP realizes the coarse-grained free energy landscape of first-principles phase stability theory~\citep{van2002effect} in a differentiable, transferable form, made practical by universal MLIPs and improved statistical-mechanical algorithms.

Our instantiation, TIP[UMA], is built on Universal Models for Atoms (UMA)~\citep{wood2025uma}, an MLIP trained across chemistries on large DFT datasets.
It is trained in three stages of increasing physical fidelity (\cref{fig:overview}d).
Pre-training fits dense quasi-harmonic free energy grids over $(T,P)$, which are computationally cheap and abundant but approximate each potential energy basin as harmonic.
Mid-training corrects toward anharmonic, absolute free energies computed from nonequilibrium MD.
Because a TIP exposes the thermodynamic responses as consistent derivatives of a single free energy surface, measured responses can supervise TIP directly, whereas an MLIP could obtain the same observables only through ensemble reweighting.
Post-training exploits this, calibrating against experiment by matching measured responses, such as heat capacity, for ordered phases and through differentiable reweighting of Monte Carlo simulations for disordered phases.

\begin{figure*}[tp]
\centering
\includegraphics[width=\textwidth]{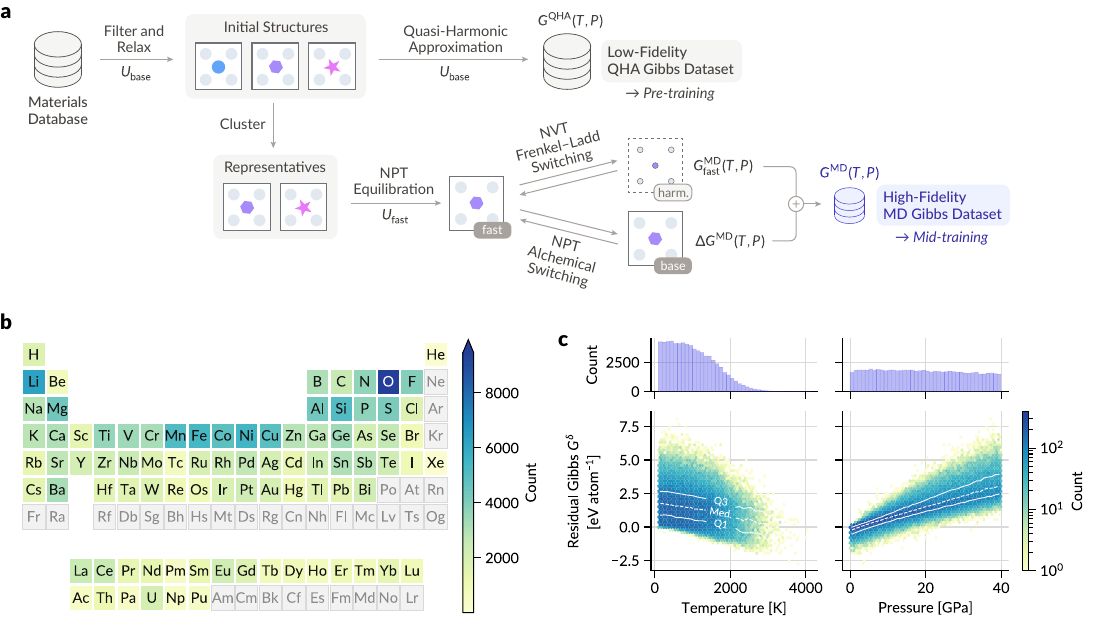}
\caption{%
\textbf{Multi-fidelity Gibbs free energy dataset.}
\textbf{(a)} Data-generation workflow.
Structures from a materials database are filtered and relaxed with the base potential $U_\mathrm{base}$ (UMA~\citep{wood2025uma}); the quasi-harmonic approximation (QHA) on each yields the low-fidelity QHA Gibbs dataset $G^\mathrm{QHA}(T,P)$.
Clustering selects representatives, which are equilibrated under $NPT$ dynamics with a fast potential $U_\mathrm{fast}$ (Orb-v3~\citep{rhodes2025orb}) and then processed by $NVT$ Frenkel--Ladd switching~\citep{frenkel1984new} (absolute free energy against a harmonic Einstein crystal reference) and $NPT$ alchemical switching (transfer from the fast to the base Hamiltonian); combining the two contributions gives the high-fidelity MD Gibbs dataset $G^\mathrm{MD}(T,P)$.
\textbf{(b)} Elemental coverage of the MD Gibbs dataset: the periodic table is colored by the number of dataset entries containing each element.
\textbf{(c)} Distribution of the residual Gibbs free energy $G^\delta$ of the MD Gibbs dataset versus temperature (left) and pressure (right), shown as two-dimensional histograms with the marginal count distributions above; white lines mark the quartiles (Q1, median, Q3).
}
\label{fig:dataset}
\end{figure*}

\subsection{Multi-Fidelity Gibbs Free Energy Dataset}

Training a structure-conditioned free energy surrogate requires thermodynamic labels at scale, which no single method can provide both efficiently and accurately.
We therefore assemble a multi-fidelity dataset that combines abundant, approximate labels with sparse but accurate labels (\cref{fig:dataset}a).
We draw the initial structures from the Materials Project~\citep{jain2013commentary}, retain metastable entries within $0.2$~eV/atom of the convex hull ($123{,}424$ structures), and relax each to an ordered branch representative $\bm{x}^\circ$ with UMA~\citep{wood2025uma}.
A representative subset is selected by BIRCH clustering~\citep{zhang1996birch} of the UMA structure embeddings, so that computationally expensive labels are assigned to a structurally diverse set of representatives while low-fidelity labels cover all structures.

The low-fidelity signal is the QHA Gibbs free energy $G^\mathrm{QHA}(T,P)$, computed for every structure; it is computationally inexpensive and includes volume-dependent harmonic vibrational contributions but neglects anharmonic mode coupling.
The high-fidelity signal is the absolute Gibbs free energy $G^\mathrm{MD}(T,P)$ from nonequilibrium MD, computed only on representatives by combining a Frenkel--Ladd absolute reference on a fast surrogate potential (Orb-v3~\citep{rhodes2025orb}, trained on the same materials dataset as UMA) with an alchemical transfer to the accurate target potential; it includes anharmonic effects but is far more demanding.
After quality filtering, the dataset comprises over $120{,}000$ QHA Gibbs free energy surfaces and roughly $95{,}000$ MD state points, spanning temperatures from $0$~K up to $4000$~K and pressures up to $40$~GPa.
The combined labels cover much of the periodic table (\cref{fig:dataset}b) and a broad range of the thermodynamic residual $G^\delta$ over temperature and pressure (\cref{fig:dataset}c).

\begin{figure}[!tbp]
\centering
\includegraphics[width=\columnwidth]{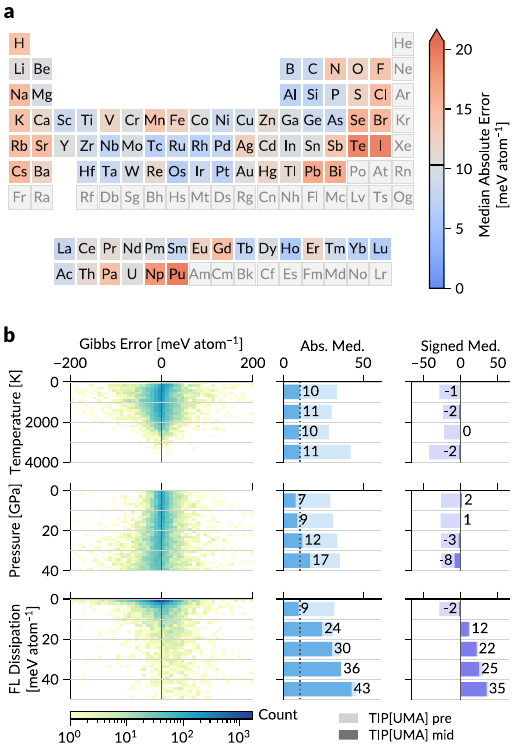}
\caption{%
\textbf{Validation error analysis.}
\textbf{(a)} Element-wise validation error.
Periodic-table map of the median absolute Gibbs error for each element, computed over the filtered validation data points that contain that element; elements represented by fewer than $20$ validation state points are shown in gray.
The black tick on the colorbar marks the global median absolute error.
\textbf{(b)} Thermodynamic and dissipation trends.
Signed Gibbs error distributions as functions of temperature, pressure, and Frenkel--Ladd (FL) dissipation for the filtered validation set.
Left panels show two-dimensional count histograms of the mid-trained model on a logarithmic color scale; right panels show the binned median absolute error and the signed median error (meV~atom$^{-1}$) for the pre-trained (light) and mid-trained (dark) models.
The dotted line in the absolute-median panels marks the global median absolute error.
}
\label{fig:validation}
\end{figure}

\subsection{Validation and Error Analysis}

After pre-training on the quasi-harmonic $(T,P)$ grids and mid-training on the anharmonic MD labels, we evaluate TIP[UMA] on a held-out validation set not used in training (\cref{fig:validation}).
Across the filtered validation set, the model attains a global median absolute Gibbs energy error of $10.3$~meV~atom$^{-1}$.
For context, the descriptor-based Gibbs free energy model of \citet{bartel2018physical} reported test errors of $46$--$60$~meV~atom$^{-1}$, although the targets and validation datasets differ.
The TIP[UMA] error is also comparable to the energy scale governing polymorph competition: experimentally observed inorganic phases have a median energy above the convex hull of approximately $15$~meV~atom$^{-1}$~\citep{sun2016thermodynamic}, while chemistry-dependent metastability windows can extend to tens or hundreds of meV~atom$^{-1}$~\citep{aykol2018thermodynamic}.
The element-resolved errors concentrate in chemically distinct regions (\cref{fig:validation}a): the halogens, chalcogens, alkali and alkaline-earth metals, heavy $p$-block elements, and actinides carry the largest median errors, whereas most transition metals sit at or below the global median.
These groups include soft, polarizable, and strongly anharmonic chemistries whose absolute free energies are difficult to converge, and rare chemistries like actinides are sparsely represented in the Materials Project pool.

Error trends with thermodynamic state and the underlying nonequilibrium MD convergence are more systematic (\cref{fig:validation}b).
Comparing the pre-trained and mid-trained models (light versus dark bars), mid-training on the MD labels lowers the median error about threefold at every temperature, correcting the offset between the quasi-harmonic pre-training and the absolute MD reference.
For the mid-trained model, the error grows only weakly with temperature and more strongly at high pressure, and its strongest dependence is on the Frenkel--Ladd switching dissipation of the high-fidelity calculation.
At the highest dissipation, the signed median error also becomes positive, indicating systematic bias rather than symmetric scatter.
Such high-dissipation labels are only a small fraction of the dataset (\cref{fig:validation}b, left).
Because the dissipated work measures how far each switching trajectory departs from reversibility, the positive bias at high dissipation reflects the convergence of the free energy labels rather than a failure of the surrogate.
The least-converged calculations are removed by a per-leg dissipation filter during curation.
These trends indicate that improving label coverage and convergence is more likely to reduce worst-case errors than increasing the model capacity.

\begin{figure*}[tp]
\centering
\includegraphics[width=\textwidth]{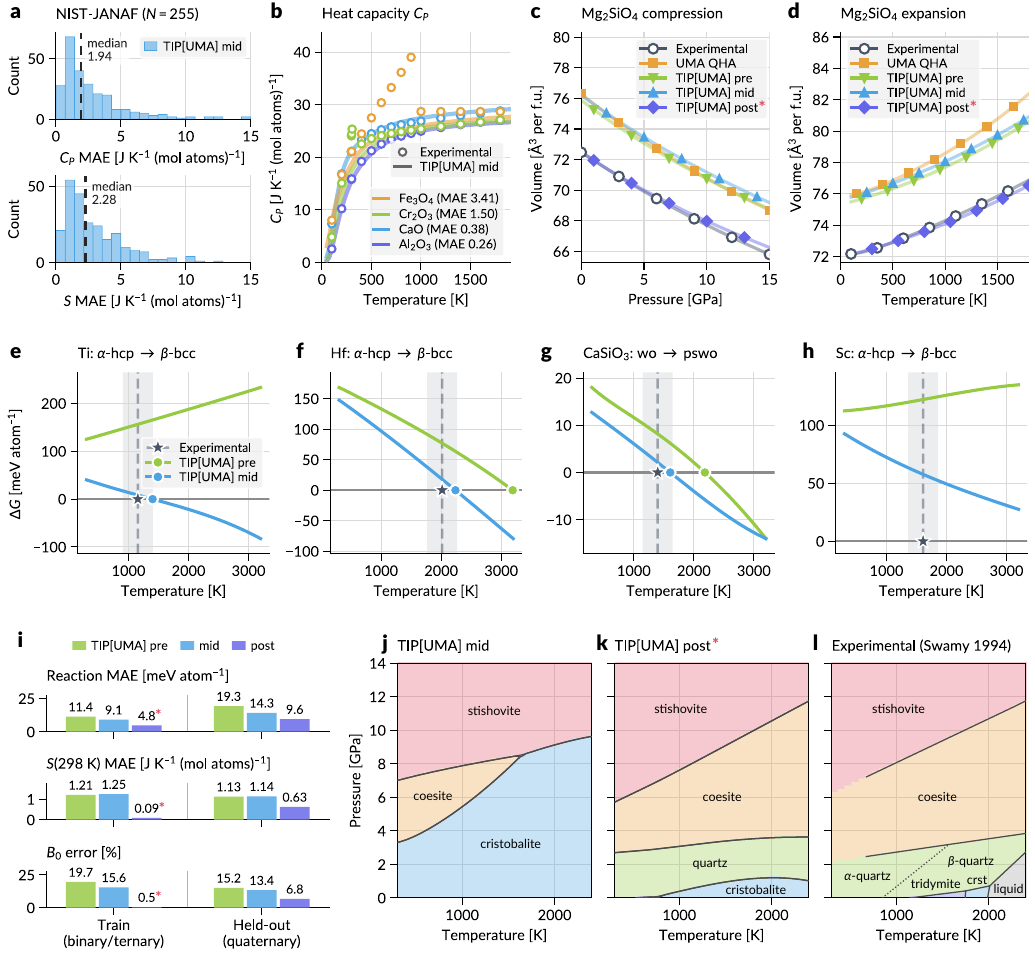}
\caption{%
\textbf{Thermodynamics of ordered crystalline phases from a learned free energy surface.}
\textbf{(a)} Mid-trained mean absolute error in heat capacity $C_P$ and standard entropy $S(298~\mathrm{K})$ across the NIST--JANAF benchmark of $255$ stoichiometric solids, with medians of $1.94$ ($C_P$) and $2.28$ ($S$)~J~K$^{-1}$~(mol\,atoms)$^{-1}$.
\textbf{(b)} Mid-trained heat capacity $C_P(T)$ (solid lines) against NIST--JANAF values~\citep{chase1998janaf} (open circles). The larger deviations for \ce{Cr2O3} and \ce{Fe3O4} occur near magnetic $\lambda$-anomalies (N\'eel and Curie transitions) outside the model's vibrational scope.
\textbf{(c)} Isothermal compression $V(P)$ at $300$~K and \textbf{(d)} thermal expansion $V(T)$ at $P \approx 0$ for forsterite \ce{Mg2SiO4}, comparing experiment (Holland--Powell~\citep{holland2011improved}), the UMA quasi-harmonic reference, and TIP[UMA] across pre-, mid-, and post-training. The quasi-harmonic reference and the pre- and mid-trained models share the known PBE volume overestimate, which post-training removes.
\textbf{(e--h)} Temperature-driven polymorphic transitions, shown as the per-atom Gibbs difference $\Delta G(T)$ between the high- and low-temperature phases for the pre-trained and mid-trained models: \textbf{(e)} \ce{Ti}, \textbf{(f)} \ce{Hf}, \textbf{(g)} \ce{CaSiO3}, and \textbf{(h)} \ce{Sc}. The $\Delta G = 0$ crossing (circle) is the predicted transition temperature, and the dashed line and star mark experimental transition~\citep{dinsdale1991sgte,richet1991casio3} with the shaded band spanning a $\pm250$~K tolerance. The pre-trained surface shows no crossing (\ce{Ti}, \ce{Sc}) or one far above experiment (\ce{Hf}, \ce{CaSiO3}), whereas mid-training recovers the transitions, \ce{Sc} only in part.
\textbf{(i)} Accuracy of the pre-, mid-, and post-trained models on the Holland--Powell ordered-oxide benchmark for the reaction Gibbs free energy, standard entropy, and bulk modulus $B_0$, on a training partition (binary and ternary oxides) and a held-out partition (quaternary oxides). Post-training improves every metric on both, including the held-out quaternaries.
\textbf{(j--l)} \ce{SiO2} pressure--temperature phase diagram for \textbf{(j)} the mid-trained model, \textbf{(k)} the post-trained model, and \textbf{(l)} the experimentally assessed topology~\citep{swamy1994thermodynamic}. The mid-trained model spuriously stabilizes cristobalite at ambient conditions, whereas post-training recovers the low-pressure quartz field and the quartz, coesite, stishovite ordering within the supplied candidate set.
Red asterisks mark comparisons for which the experimental value was included in calibration.
}
\label{fig:ordered}
\end{figure*}

\subsection{Ordered Crystalline Phases}

For an ordered crystalline branch, the equilibrium response functions are derivatives of the Gibbs free energy, and TIP[UMA] predicts this surface from a single relaxed structure.
From this, we evaluate heat capacities, equations of state, polymorphic transitions, and a phase diagram (\cref{fig:ordered}).

For $255$ stoichiometric solids curated from the NIST--JANAF thermochemical tables~\citep{chase1998janaf} (Methods), the mid-trained model predicts the standard entropy and heat capacity with median absolute errors of $2.28$ and $1.94$~J~K$^{-1}$~(mol\,atoms)$^{-1}$ (\cref{fig:ordered}a).
The predicted per-atom heat capacity tracks the NIST--JANAF values~\citep{chase1998janaf} closely for \ce{Al2O3} and \ce{CaO} over the full temperature range, correctly recovering $\hat{C}_P \to 0$ as $T \to 0$ (\cref{fig:ordered}b).
Because the MD labels do not directly constrain heat capacity, mid-training regularizes $C_P$ toward the frozen quasi-harmonic model, while the analytic Einstein term preserves the low-temperature limit.
Mid-training adds an anharmonic free energy correction without replacing the quasi-harmonic low-temperature heat-capacity behavior with the classical MD limit.
The larger errors for \ce{Cr2O3} and \ce{Fe3O4} arise from magnetic $\lambda$-anomalies (the antiferromagnetic N\'eel and ferrimagnetic Curie transitions) in the measured $C_P$, which lie outside the vibrational contributions.
An additional explicit magnetic or electronic term could capture these contributions (see Discussion).
The volumetric response of forsterite \ce{Mg2SiO4} shows that the predicted compression and thermal expansion reproduce the UMA reference behavior (\cref{fig:ordered}c,d).
Both the UMA QHA reference and the model predictions retain the known volume overestimate of the Perdew--Burke--Ernzerhof (PBE) functional, inherited from its systematic underbinding and a property of the training target rather than the adapter.
Post-training against experimental labels later removes it (see below).

We next examine temperature-driven polymorphic transitions involving high-temperature phases that the quasi-harmonic surface does not describe adequately (\cref{fig:ordered}e--h).
These transitions are entropy-driven and require anharmonic treatment: the high-temperature bcc metals are dynamically unstable at $0$~K and have no harmonic reference, while for \ce{CaSiO3} a harmonic reference exists but omits the anharmonic entropy that sets the transition~\citep{souvatzis2008entropy,hellman2013temperature,thomas2013finite,van2015free}.
Consistent with these limitations, the pre-trained model produces no $\Delta G = 0$ crossing for \ce{Ti} or \ce{Sc}, and for \ce{Hf} and \ce{CaSiO3} it crosses only far above experiment (near $3200$ and $2200$~K, against $2016$ and $1398$~K).
After mid-training on nonequilibrium MD free energies, the model predicts three of the four (\ce{Sc} only in part, below), placing the \ce{Ti} and \ce{Hf} $\alpha$-hcp $\to$ $\beta$-bcc transitions and the \ce{CaSiO3} wollastonite $\to$ pseudowollastonite transition within ${\sim}250$~K of experiment.
Of the four, only \ce{Ti} has both phases in the MD training set, while \ce{Sc} and \ce{Hf} have their high-temperature bcc phase held out and \ce{CaSiO3} has neither polymorph.
These entropy-driven transitions are recovered even though mid-training updates only a lightweight adapter on a frozen encoder, indicating that the MD labels provide transferable anharmonic corrections.
The \ce{Sc} $\alpha$-hcp $\to$ $\beta$-bcc transition (\cref{fig:ordered}h) is recovered only in part: mid-training moves $\Delta G$ toward a crossing but does not produce a crossing at the measured transition temperature, showing that the anharmonic correction remains imperfect where high-fidelity training data are sparse.

The pre- and mid-trained models so far inherit the systematic volume overestimate of the PBE reference.
Post-training (\cref{fig:overview}d) removes this bias, calibrating the model against the Holland--Powell experimental thermodynamic assessment over the Si--Al--Mg--Ca--O chemical space~\citep{holland2011improved}.
The model is fitted on the binary and ternary oxides, with the quaternary oxides held out (Methods and \siref{D}).
The calibration lowers the bulk modulus error against experiment from ${\sim}13\%$ to a few percent, decreases the reaction Gibbs free energy and entropy errors, and removes the ${\sim}5\%$ volume overestimate from the PBE offset (\cref{fig:ordered}c,d, TIP[UMA] post).
The metrics on the held-out quaternary phases improve alongside those on the fitted binary and ternary ones (\cref{fig:ordered}i), indicating that the representation learned through pre- and mid-training supports transferability, so calibration on a small experimental set refines the surface rather than fitting each chemistry independently.
The same calibration also corrects the \ce{SiO2} pressure--temperature phase diagram (\cref{fig:ordered}j--l): the mid-trained model spuriously stabilizes cristobalite at ambient conditions, and post-training restores quartz as the ambient ground state and the quartz $\to$ coesite $\to$ stishovite ordering within the supplied candidate set~\citep{swamy1994thermodynamic}.

\begin{figure*}[tp]
\centering
\includegraphics[width=\textwidth]{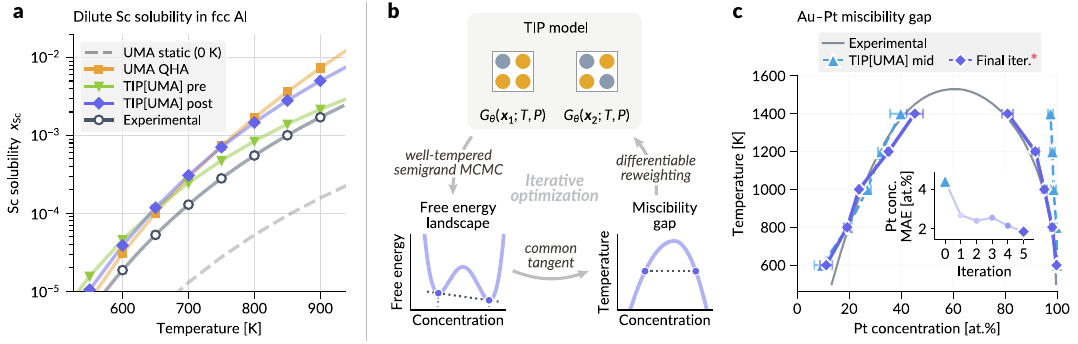}
\caption{%
\textbf{Calibration of TIP[UMA] for disordered phases using computational and experimental references.}
Lightweight fine-tuning fits the frozen free energy surrogate to a chosen reference and extends it from ordered branches to configurationally disordered solid solutions, preserving thermodynamic consistency.
\textbf{(a)} Dilute Sc solubility in fcc \ce{Al} versus temperature, obtained from the dissolution free energy relative to the adjacent \ce{Al}--\ce{Al3Sc} convex hull (\cref{sec:methods}).
By accounting for vibrational entropy, TIP[UMA] predicts the measured order of magnitude and tracks the quasi-harmonic UMA reference (UMA QHA), whereas the $0$~K static estimate falls about an order of magnitude below the measured value; \ce{Al}--\ce{Sc}-focused fine-tuning (pre to post) moves the model toward that reference.
The computed reference itself lies above the measured solubility, an offset set by the reference thermodynamics rather than the surrogate.
\textbf{(b)} Workflow for a disordered phase: TIP[UMA] scores lattice configurations, well-tempered semigrand Monte Carlo samples the mixing free energy, a common-tangent construction yields the miscibility gap, and differentiable reweighting calibrates the surrogate.
\textbf{(c)} \ce{Au}--\ce{Pt} miscibility gap, composition in \ce{Pt} atomic percent (at.\%): iterated differentiable reweighting moves the TIP[UMA] binodal from its initial prediction to the experimental solvus, the Pt-concentration error falling over the iterations (inset).
Horizontal bars are the seed spread, one standard deviation of the branch Pt concentration across four independent well-tempered metadynamics seeds.
Red asterisks mark comparisons for which the experimental value was included in calibration.
}
\label{fig:calibration}
\end{figure*}

\subsection{Disordered Solid Solutions}

Ordered branches are only part of the phase stability problem: alloy solubility limits and miscibility gaps require configurational disorder in addition to ordered-branch thermodynamics.
TIP treats these systems by evaluating branch free energies across composition on a parent lattice, which defines the configurational Hamiltonian used to sample the disordered state.
That free energy can then be calibrated against a chosen reference in either of two complementary directions: \emph{bottom-up}, distilling a higher-resolution calculation for a target chemistry into the surrogate, or \emph{top-down}, matching measured phase boundaries.

The dilute solubility of Sc in fcc \ce{Al}, which depends on the vibrational entropy of dissolution in addition to configurational mixing~\citep{ozolins2001large}, provides a bottom-up example of chemistry-specific fine-tuning across disordered compositions and configurations (\cref{fig:calibration}a).
At each temperature and $P = 0$, TIP[UMA] evaluates the Gibbs free energies required to compute the dissolution free energy $\Delta G_\mathrm{sol}(T)$ relative to the adjacent \ce{Al}--\ce{Al3Sc} convex hull (\cref{eq:methods-alsc-diss}).
The equilibrium solubility then follows from the ideal-dilute relation $x_\mathrm{Sc}\approx\exp(-\Delta G_\mathrm{sol}/k_\mathrm{B}T)$ (\cref{eq:methods-alsc-solub}).
Using only $0$~K static energies from the baseline UMA potential (no vibrational free energy), the estimate increases with temperature through the ideal-mixing term but remains about an order of magnitude below the measured solubility across $600$--$900$~K.
Including branch vibrational free energies increases $x_\mathrm{Sc}$ by one to nearly two orders of magnitude over that static line, reaching the measured range even from the pre-trained model with no \ce{Al}--\ce{Sc}-specific tuning.
We then test chemistry-specific distillation by fine-tuning the surrogate on an additional set of UMA quasi-harmonic free energies for diverse fcc \ce{Al}--\ce{Sc} structures spanning compositions and configurations using a lightweight low-rank adaptation (LoRA)~\citep{hu2022lora} (Methods and \siref{B}).
The fine-tuned model then more closely reproduces the dedicated \ce{Al}--\ce{Sc} quasi-harmonic reference.
While the reference overpredicts the measured high-temperature solubility by a factor of three to four, this example evaluates fidelity to the computational reference rather than agreement with experiment.

The \ce{Au}--\ce{Pt} miscibility gap provides a top-down example in which the model is calibrated to the experimental solvus~\citep{darling1952aupt,munster1960aupt}.
Since TIP[UMA] assigns each lattice configuration a branch-conditioned Gibbs free energy that includes explicit temperature and pressure dependence, it serves as a temperature- and pressure-dependent configurational Hamiltonian analogous to a cluster expansion.
Semigrand canonical Monte Carlo~\citep{sadigh2012scalable}, accelerated by well-tempered metadynamics~\citep{barducci2008well}, therefore samples the mixing free energy across composition directly on the surrogate.
A common-tangent construction then yields the miscibility gap (\cref{fig:calibration}b), and differentiable reweighting~\citep{thaler2021learning} of the stored samples calibrates the surrogate toward the measured solvus over successive cycles.
To reduce computational cost, the structure is fully relaxed every few steps, the stored samples are reused within each cycle, and the metadynamics bias is warm-started between cycles (Methods and \siref{B}).
Over these cycles, the predicted binodal moves toward the measured solvus (\cref{fig:calibration}c), and the error in \ce{Pt} concentration falls (inset).
The same surrogate supports both ordered and disordered thermodynamics and can be specialized against computational or experimental references.
In both cases, differentiable calibration refines an effective free energy model against the available reference without fitting a separate configurational Hamiltonian, providing a flexible realization of the coarse-graining program of first-principles phase stability theory~\citep{van2002effect}.

\begin{figure*}[tp]
\centering
\includegraphics[width=\textwidth]{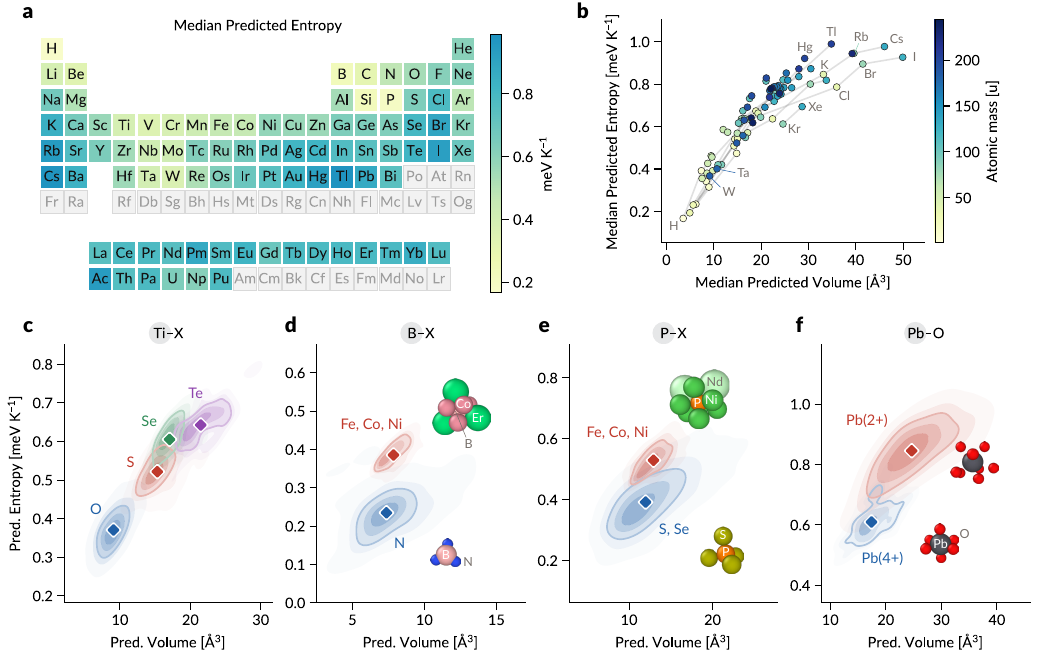}
\caption{%
\textbf{Chemical trends in predicted entropy and volume.}
Predicted per-atom entropy $\hat S$ and volume $\hat V$ from TIP[UMA] at $T=1000$~K and $P=0$~GPa, across the filtered high-fidelity dataset.
\textbf{(a)} Periodic table colored by each element's median predicted entropy.
\textbf{(b)} Element-level median entropy versus median volume, colored by atomic mass, with faint guide lines connecting elements within the same group; the two are strongly correlated.
\textbf{(c--f)} Site-resolved entropy--volume density maps, each grouping sites by one aspect of local chemistry over all sites of the central element, with diamonds marking the group medians.
\textbf{(c)} Neighbor identity, \ce{Ti}--X, for neighbors \ce{O}, \ce{S}, \ce{Se}, and \ce{Te}.
\textbf{(d,e)} Bond character: covalent \ce{B}--\ce{N} and \ce{P}--\ce{S}/\ce{P}--\ce{Se} sites versus metallic \ce{Fe}, \ce{Co}, and \ce{Ni} borides and phosphides.
\textbf{(f)} Formal valence, \ce{Pb}--\ce{O}: compact \ce{Pb^4+} versus expanded \ce{Pb^2+}, softened by a stereochemically active $6s^2$ lone pair.
}
\label{fig:trends}
\end{figure*}

\subsection{Chemical Trends in Predicted Responses}

To examine what chemical information its learned atomic contributions encode, we analyze its per-atom contributions.
Although it is supervised only on structure-level thermodynamics, its head assembles each Gibbs free energy surface additively from per-atom contributions.
This decomposition defines per-atom entropy $\hat S_i$ and volume $\hat V_i$ that sum to the structure totals ($\hat S=\sum_i \hat S_i$, $\hat V=\sum_i \hat V_i$).
We therefore test whether these learned descriptors follow established chemical trends (\cref{fig:trends}).

Across the periodic table, the predicted per-atom entropy $\hat S_i$ is largest for large, soft, polarizable elements such as the heavy alkali metals, Tl, Hg, and the heavy halogens, and smallest for compact covalent and stiff refractory species such as B, C, Si, P, and the early-to-mid transition metals (\cref{fig:trends}a).
The element-level medians of entropy and volume are strongly correlated, and both generally increase with atomic mass (\cref{fig:trends}b).
Notable exceptions are Ta and W, which remain compact with low entropy because of their strong bonding, in contrast to softer elements such as I, Cs, Tl, and Hg with similar masses, which occupy larger volumes with higher entropy.
Compared with the composition-level descriptor of the Gibbs free energy from Bartel et al.~\citep{bartel2018physical}, in which atomic mass enters directly through a reduced-mass term alongside volume, TIP[UMA] shows a related but more mixed dependence, with mass dependence mediated mainly by the bonding environment and atomic volume.
This pattern is consistent with vibrational entropy increasing as local bonding softens and effective atomic volume grows.

Within an element, much of the variation arises from local environment, and because TIP[UMA] is structure-conditioned and per-atom, it resolves variation that a composition-level descriptor averages away.
We group sites by neighbor identity, bond character, and formal valence (\cref{fig:trends}c--f).
For \ce{Ti}, predicted volume and entropy increase monotonically as the dominant neighbor becomes heavier and more polarizable from \ce{O} through \ce{S}, \ce{Se}, and \ce{Te} (\cref{fig:trends}c).
Bond character separates sites at similar predicted volume: stiff, low-coordination covalent sites, such as threefold \ce{B}--\ce{N} or \ce{P} bound to \ce{S} and \ce{Se}, fall well below the soft, high-coordination metallic sites of the corresponding borides and transition-metal phosphides (\cref{fig:trends}d,e).
This separation is consistent with the established microscopic picture of vibrational entropy: short, stiff, high-frequency bonds carry little entropy, while soft, low-frequency metallic bonding carries much more~\citep{van2002effect,garbulsky1996contribution}.
Formal valence provides an additional distinction: even when the neighboring chemistry is the same, the response descriptors separate by oxidation state.
In \ce{Pb}--\ce{O}, compact, near-octahedral \ce{Pb^4+} has lower predicted entropy, while \ce{Pb^2+} has larger predicted volume and entropy, consistent with its stereochemically active $6s^2$ lone pair (\cref{fig:trends}f).

Because only the structure-level properties are supervised and atomic decompositions of bulk responses are non-unique, $\hat S_i$ and $\hat V_i$ should be interpreted as learned descriptors rather than physical observables.
Even so, their agreement with several established trends in local chemistry provides qualitative evidence that the representation captures chemically meaningful structure.

\section{Discussion}
\label{sec:discussion}

TIP predicts a branch-conditioned Gibbs free energy surface $G(\bm{x}^\circ;\,T,P)$ from a relaxed crystal structure.
Automatic differentiation yields volume, entropy, and heat capacity from that surface rather than from separate predictions.
The model is trained across fidelities, from quasi-harmonic grids to molecular dynamics free energies, and calibrated against either higher-resolution calculations or experiment.
TIP therefore links several approaches for first-principles thermodynamics within one differentiable model.

The present model includes vibrational and configurational contributions but omits magnetic and electronic entropy.
The heat capacity anomalies of \ce{Cr2O3} and \ce{Fe3O4} (\cref{fig:ordered}) illustrate this limitation.
Including this entropy is a prerequisite for magnetic and mixed-valence solids, where it can qualitatively rearrange the phase diagram~\citep{zhou2006configurational}.
The modular adapter could include these additional residual terms analogous to magnetic models in CALPHAD~\citep{hillert1978model}.
Alternatively, charge- and spin-informed per-atom features, for example from CHGNet~\citep{deng2023chgnet}, could be supplied to the encoder.
This would offer a complementary route in which the model resolves the magnetic and electronic entropy from electronic-structure-derived features directly.
Point defects~\citep{mosquera2023imperfections}, surfaces~\citep{du2023surface}, and interfaces lie outside the ordered bulk branches.
Their free energies scale per defect or per area rather than with bulk atom count.
Extending TIP to these objects would require chemical potentials and finite-size or slab treatments.
Occupational disorder is resolved here by conditioning each branch on a specific ordering and sampling compositions with semigrand canonical Monte Carlo, so the configurational search remains computationally costly.
Supplying site occupancies directly as continuous inputs~\citep{nam2025alchemical} could reduce or replace explicit configurational sampling and enable efficient alloy thermodynamics.

A second limitation is the quality of the high-fidelity reference.
TIP[UMA]'s worst-case error tracks the Frenkel--Ladd switching dissipation of the underlying calculation more strongly than with the tested variables (\cref{fig:validation}).
Accuracy is therefore set by the quality of the free energy labels, making label convergence and coverage major determinants of accuracy.
Improvements should therefore prioritize tighter reversibility of the switching simulations, broader chemical and structural coverage, and extension to higher pressures and disordered structures.
The underlying level of theory imposes a separate limit: the PBE underbinding inherited from UMA sets a systematic volume and bulk modulus bias that experimental post-training can correct empirically but cannot remove from the underlying reference.
Rebuilding the same construction on a base potential trained at a higher level of theory would address this bias at its source.
Finally, extension to liquids requires both new absolute reference states~\citep{leite2016uhlenbeck} and input representations that do not depend on a relaxed ordered branch.
Meeting both requirements would enable full solid--liquid phase diagrams.

Together, these extensions define a broader goal: a differentiable thermodynamic foundation model calibrated to first-principles and experimental data and evaluated at MLIP-like inference compute cost.
Such a model would make finite-temperature free energies and response functions routine inputs to materials discovery.

\section{Methods}
\label{sec:methods}

\subsection{Branch-conditioned Gibbs free energy}

TIP models the Gibbs free energy of an ordered branch labeled by its relaxed reference structure $\bm{x}^\circ$.
Each $\bm{x}^\circ$ is obtained by a symmetry-constrained relaxation of the static potential energy surface.
The branch-conditioned isothermal-isobaric ($NPT$) partition function restricts configuration space to the basin $\mathcal{B}(\bm{x}^\circ)$ that relaxes onto $\bm{x}^\circ$, and the resulting $G(\bm{x}^\circ;\,T,P)$ marginalizes vibrational and cell fluctuations within the branch.
The formal branch map, the partition function construction, and the three theoretical assumptions (branch representability, finite-$(T,P)$ branch stability, and fixed-branch scope) that make this target well defined are given in \siref{A}.

For modeling, we use the static/thermodynamic split of~\cref{eq:G-split-main}, with $U^\circ := U(\bm{x}^\circ)$ the static energy of the branch representative and $\hat{G}^\delta$ absorbing all finite-$(T,P)$ contributions.
The zero-point energy is excluded from the labels (\cref{eq:methods-g0}), to maintain consistency with the classical MD reference, which carries no zero-point contribution.
This convention also avoids an ill-defined harmonic zero-point energy for dynamically unstable branches with imaginary modes.
First-order equilibrium responses follow as branch-conditioned thermodynamic derivatives,
\begin{equation}
\label{eq:methods-S-V}
    S = -\left(\frac{\partial G}{\partial T}\right)_{\!P}, \qquad V = \left(\frac{\partial G}{\partial P}\right)_{\!T},
\end{equation}
and the second-order responses (heat capacity, isothermal bulk modulus, and volumetric thermal expansion) follow from higher derivatives of the same $G$,
\begin{align}
\label{eq:methods-second-order}
    C_P &= -T\left(\frac{\partial^2 G}{\partial T^2}\right)_{\!P}, \\
    B &= -V\left(\frac{\partial^2 G}{\partial P^2}\right)_{\!T}^{-1}, \label{eq:methods-BT}\\
    \alpha &= \frac{1}{V}\frac{\partial^2 G}{\partial T\, \partial P}.
\end{align}

\subsection{Gibbs free energy model}

We parametrize the thermodynamic residual atomwise to respect extensivity,
\begin{equation}
\label{eq:methods-Gd-atomwise}
    \hat{G}^\delta(\bm{x}^\circ;\,T,P) = \sum_{i=1}^N \hat{g}^\delta_i(\bm{h}_i;\,T,P),
\end{equation}
where $\bm{h}_i$ is the per-atom latent feature produced by the thermodynamic backbone introduced in the main text (\cref{fig:overview}c).
The backbone is an equivariant PaiNN encoder~\citep{schutt2021equivariant} that takes the frozen MLIP's scalar ($l{=}0$) and vector ($l{=}1$) per-atom features, and refines them by equivariant message passing over the local geometry.
The encoder receives neither temperature nor pressure, so $(T,P)$ enter only at the head; we use a trunk with $r_\mathrm{max}=6$~\AA, $d=64$ latent features, $16$ radial basis functions, and $3$ message-passing layers.
Each per-atom residual is written so that the volume is exactly the pressure derivative,
\begin{equation}
\label{eq:methods-gd-decomp}
    \hat{g}^\delta_i(\bm{h}_i;\,T,P) = \hat{g}^{(0)}_i(\bm{h}_i;\,T) + \int_0^P \hat{V}_i(\bm{h}_i;\,T,p)\,\mathrm{d}p,
\end{equation}
so that $\hat{V} = \partial \hat{G}^\delta/\partial P = \sum_i \hat{V}_i$ holds by construction, while $\hat{S} = -\partial \hat{G}/\partial T$ follows by automatic differentiation in temperature.

The zero-pressure reference branch combines an Einstein oscillator vibrational term with a low-order polynomial in a normalized temperature $\tau := T/T_\mathrm{max}$ (with $T_\mathrm{max} = 4000$~K a fixed global scale),
\begin{multline}
\label{eq:methods-g0}
    \hat{g}^{(0)}_i(\bm{h}_i;\,T) = c_0 + c_1\tau + c_2\tau^2 + c_3\tau^3 \\
    + 3 k_\mathrm{B} T\ln\!\big(1-e^{-\theta_\mathrm{E}/T}\big),
\end{multline}
with the coefficients $\{c_k\}$ and the Einstein temperature $\theta_\mathrm{E} = T_\mathrm{max}[\mathrm{softplus}(\tilde{\theta}) + \theta_{\min}]$ predicted from $\bm{h}_i$.
This five-coefficient form ($c_0$--$c_3$ and $\theta_\mathrm{E}$) is the model used in pre-training.
Mid-training augments this branch with two gated anharmonic terms, adding the coefficients $c_\mathrm{log}$ and $c_\mathrm{inv}$ for seven in total,
\begin{equation}
\label{eq:methods-g0-aug}
    \hat{g}^{(0),\mathrm{mid}}_i(\bm{h}_i;\,T) = \hat{g}^{(0)}_i(\bm{h}_i;\,T) + w(\tau)\!\left[c_\mathrm{log}\,\tau\ln\tau + \frac{c_\mathrm{inv}}{\tau}\right],
\end{equation}
where sigmoidal gate $w(\tau) = (\tau/\tau_\mathrm{E})^{p}/[1+(\tau/\tau_\mathrm{E})^{p}]$, with $\tau_\mathrm{E} := \theta_\mathrm{E}/T_\mathrm{max}$ and $p=4$.
This confines the $\tau\ln\tau$ and $1/\tau$ corrections to temperatures above $\theta_\mathrm{E}$, so they represent anharmonic behavior beyond the quasi-harmonic reference while leaving the low-temperature branch intact.
Each atomic volume contribution follows a Murnaghan-like pressure response~\citep{murnaghan1944compressibility} whose per-atom parameters are softplus-transformed quadratic functions of $\tau$,
\begin{align}
\label{eq:methods-eos-params}
    V_{0,i}(\tau) &= \mathrm{softplus}\big(\textstyle\sum_{k=0}^{2} a^{V_0}_k\tau^k\big) + V_{0,\min}, \\
    B_{0,i}(\tau) &= \mathrm{softplus}\big(\textstyle\sum_{k=0}^{2} a^{B_0}_k\tau^k + B_\mathrm{off}\big) + B_{0,\min}, \label{eq:methods-B0}\\
    B'_{0,i}(\tau) &= 1 + \mathrm{softplus}\big(\textstyle\sum_{k=0}^{2} a^{B'_0}_k\tau^k\big) + B'_{0,\min},
\end{align}
the additive floors enforce $V_{0,i}>0$, $B_{0,i}>0$, $B'_{0,i}>1$, and $\theta_\mathrm{E}>0$.
An offset initializes the initial bulk modulus near $B_\mathrm{off}=100$~GPa.
This parametrization gives the closed-form atomwise volume and analytic pressure integral
\begin{align}
\label{eq:methods-V}
    \hat{V}_i(\tau,P) &= V_{0,i}(\tau)\,x_i^{-1/B'_{0,i}(\tau)}, \\
\label{eq:methods-pint}
    \int_0^P \! \hat{V}_i(\tau,p)\,\mathrm{d}p &= \frac{V_{0,i}(\tau)\,B_{0,i}(\tau)}{B'_{0,i}(\tau)-1}\big(x_i^{(B'_{0,i}(\tau)-1)/B'_{0,i}(\tau)}-1\big),
\end{align}
with $x_i := 1 + B'_{0,i}(\tau)\,P/B_{0,i}(\tau)$.
The head uses the Murnaghan-like form because its pressure integral~\cref{eq:methods-pint} is closed-form and yields an analytic Gibbs free energy, whereas the Vinet form used for the quasi-harmonic fit~\citep{vinet1987temperature} requires a numerical root solve.
An ablation comparing the two forms gave comparable accuracy.

\subsection{Multi-fidelity dataset}

\textit{Source curation and splits.}
The Materials Project pool~\citep{jain2013commentary} is filtered to metastable entries within $0.2$~eV/atom of the convex hull ($123{,}424$ structures) and each structure is relaxed with UMA (\texttt{UMA-S-1.1}, \texttt{omat} task)~\citep{wood2025uma} to its static branch representative via ASE~\citep{hjorth2017atomic}.
A representative subset of $26{,}226$ structures is obtained by BIRCH clustering~\citep{zhang1996birch} of the UMA embeddings under a Euclidean radius threshold of $0.15$, split $90/10$ into $23{,}603$ training and $2{,}623$ validation representatives.
The remaining structures are mapped to the nearest representative in embedding space, and the low-fidelity stream is sampled with inverse-cluster-size weights.

\textit{Low-fidelity quasi-harmonic labels.}
The low-fidelity signal is the quasi-harmonic approximation~\citep{baroni2001phonons,togo2015first}.
At each cell volume, the Helmholtz free energy is the sum of the quantum harmonic oscillator free energies over the Brillouin-zone phonon branches.
Imaginary-frequency modes, whose harmonic free energy is ill-defined, are excluded from this sum.
Phonons are computed by finite displacements of $0.01$~\AA{} in an approximately isotropic supercell of minimum side length $12$~\AA{} with Phonopy~\citep{togo2015first}.
Thermal properties are then evaluated from $0$~K to $T_m+500~\mathrm{K}$ in $5$~K steps, where $T_m$ is a predicted melting point~\citep{hong2022melting} read from the Materials Project metadata.
Repeating this over a $12$-point isotropic strain grid ($\eta\in[-0.08,0.03]$) traces the free energy against volume.
A Legendre transform at fixed pressure and a Vinet equation-of-state fit~\citep{vinet1987temperature} then yield the branch-conditioned Gibbs free energy, with the zero-point contribution subtracted.
The governing equations are given in \siref{B}.
For training, one $(T,P)$ point is sampled per structure, with $T$ drawn uniformly over the same $0$ to $T_m+500$~K range used for the grid and $P\sim\mathcal{U}([0,40]~\mathrm{GPa})$.
Samples with equation-of-state RMS residual above $5$~meV/atom or a negative fitted bulk modulus are rejected, leaving $122{,}913$ accepted surfaces ($111{,}402$ training / $11{,}511$ validation).

\textit{High-fidelity molecular dynamics labels.}
The high-fidelity signal is the absolute Gibbs free energy from nonequilibrium MD, computed only on representatives.
Each representative is relaxed with a fast, non-equivariant Orb-v3 potential (\texttt{orb-v3-conservative-inf-omat})~\citep{rhodes2025orb} and sampled at several independent state points (${\sim}4$ per material on average), each an independent draw of $T\sim\mathcal{U}([100~\mathrm{K},\,T_m+500~\mathrm{K}])$ and $P\sim\mathcal{U}([0,40]~\mathrm{GPa})$.
The absolute Gibbs free energy is assembled as
\begin{equation}
\label{eq:methods-Ghf}
    G^\mathrm{UMA} = G_\mathrm{FL}^\mathrm{Orb} + \Delta G^{\mathrm{Orb}\to\mathrm{UMA}},
\end{equation}
a Frenkel--Ladd reference on the Orb-v3 Hamiltonian followed by an alchemical transfer to the equivariant UMA target at the same $(T,P)$, obtained in the three steps below.

\textit{Step 1: $NPT$ equilibration.}
At each state point, an approximately isotropic ${\sim}1{,}000$-atom supercell is equilibrated under $NPT$ for ${\sim}10$~ps with a $1$~fs timestep, an isotropic Martyna--Tobias--Klein barostat~\citep{martyna1994constant}, and a Nos\'e--Hoover chain thermostat~\citep{martyna1992nose} ($100$~fs coupling).

\textit{Step 2: Frenkel--Ladd switching.}
The absolute Helmholtz free energy of the Orb-v3 crystal is obtained by Frenkel--Ladd thermodynamic integration~\citep{frenkel1984new} at the fixed equilibrated cell ($NVT$, Langevin thermostat).
A mixed Hamiltonian $U(\bm{r};\lambda)=(1-\lambda)\,U(\bm{r})+\lambda\,U_\mathrm{harm}(\bm{r})$ connects the physical potential to an Einstein crystal reference $U_\mathrm{harm}$, whose per-atom spring constants $k_i = 3k_\mathrm{B}T/\langle\Delta r_i^2\rangle$ are estimated from a $10$~ps mean-squared-displacement run.
The coupling is then driven along a smooth polynomial schedule $\lambda(t)$ with zero slope at both endpoints~\citep{deKoning1996einstein,nam2025alchemical}: forward over $25$~ps, re-equilibrated in the harmonic reference for $5$~ps, then backward over $25$~ps.
Symmetrizing the accumulated nonequilibrium work $W^s$ over the forward and backward legs~\citep{de1999optimized} cancels the leading dissipation,
\begin{equation}
\label{eq:methods-dF}
    \Delta F = \tfrac{1}{2}\big(\mathbb{E}_{\mathbb{P}_{0\to1}}[W^s] - \mathbb{E}_{\mathbb{P}_{1\to0}}[W^s]\big),
\end{equation}
and the absolute branch-conditioned Gibbs free energy follows by adding the closed-form Einstein reference $F_\mathrm{harm}$, the $PV$ term at the equilibrated cell, and a center-of-mass correction $F_\mathrm{COM}$,
\begin{equation}
\label{eq:methods-Gfl}
    G_\mathrm{FL}(\bm{x}^\circ;\,T,P) = F_\mathrm{harm}(T) - \Delta F + PV + F_\mathrm{COM}(T).
\end{equation}
The reference free energy and the center-of-mass correction are given in \siref{B}.

\textit{Step 3: Alchemical switching.}
To avoid a full Frenkel--Ladd calculation on the computationally expensive UMA potential, we transfer the absolute free energy from the Orb-v3 Hamiltonian to UMA at the same state $(\bm{x}^\circ;\,T,P)$.
The mixed Hamiltonian $U(\bm{r};\lambda)=(1-\lambda)\,U^\mathrm{Orb}+\lambda\,U^\mathrm{UMA}$ is driven under $NPT$ along the same schedule $\lambda(t)$: forward over $5$~ps, re-equilibrated at UMA for $1$~ps, then backward over $5$~ps.
Symmetrizing the two legs as in~\cref{eq:methods-dF} yields the Gibbs free energy difference $\Delta G^{\mathrm{Orb}\to\mathrm{UMA}}$, and the absolute free energy transfers by the triangle relation of~\cref{eq:methods-Ghf}.
Because Orb-v3 and UMA are trained on overlapping density functional theory data, this single alchemical step is a small perturbation. The general two-Hamiltonian derivation is given in \siref{B}.

\textit{Quality filtering and dataset.}
Each state point must pass four filters that reject unconverged or non-crystalline runs.
The equilibrated cell edge must stay within a factor of two of its $0$~K value (scale factor in $[0.5,2.0]$), excluding collapse or sublimation.
The atoms' time-averaged displacement from their lattice sites must not exceed $1.0$~\AA, removing configurations that melt or reconstruct.
The $NPT$ cell volume must reach equilibrium within $8{,}000$ of its $10{,}000$ steps, leaving a sufficient production window.
The residual switching dissipation, half the sum of the forward and backward nonequilibrium work, must stay below $0.05$~eV/atom for both the Frenkel--Ladd and alchemical switches.
After filtering, $95{,}308$ state points remain ($85{,}774$ training / $9{,}534$ validation; $22{,}894$ / $2{,}548$ unique materials).
Complete dataset sizes are tabulated in \siref{C}.
Because $T_m$ only sets the sampling ceiling, state points above it are retained as the metastable continuation of the crystalline branch.

\subsection{Multi-fidelity training}

\textit{Pre-training.}
Both training stages supervise the thermodynamic residual $G^\delta = G - U^\circ$ rather than the absolute $G$.
Pre-training trains the model from scratch with a graph-normalized $L_1$ loss on the residual and its differentiated responses,
\begin{multline}
\label{eq:methods-loss-low}
    \mathcal{L}_\mathrm{pre} = \mathcal{L}_G + \lambda_V\,\mathcal{L}_V + \lambda_S\,\mathcal{L}_S \\
    + \lambda_{C_P}\,\mathcal{L}_{C_P} + \lambda_B\,\mathcal{L}_B + \lambda_\alpha\,\mathcal{L}_\alpha,
\end{multline}
where $\mathcal{L}_G$ matches the QHA Gibbs residual, and the remaining terms match its volume, entropy, heat capacity, isothermal bulk modulus, and thermal expansion.
Hence, the full first- and second-order $(T,P)$ structure of $G$ is supervised.
The response labels come from the QHA equation-of-state fit and finite-temperature stencils.
Extensive terms are normalized per atom.
Per-sample response losses are capped to prevent rare pathological states from dominating optimization (loss weights in \siref{C}).
Optimization uses Adam~\citep{kingma2014adam} (learning rate $6\times10^{-4}$, per-GPU batch size $32$, gradient clipping at norm $10$) over ${\sim}125{,}000$ optimization steps.

\textit{Mid-training.}
Mid-training initializes from the pre-trained checkpoint and applies a LoRA update~\citep{hu2022lora} to the linear layers of the thermodynamic backbone and head, with all other weights frozen.
The update has rank $8$, scaling factor $16$, and no dropout, adding ${\sim}44{,}000$ trainable parameters.
Mid-training activates the augmented Einstein basis of~\cref{eq:methods-g0-aug} by adding two anharmonic coefficients initialized to zero.
The training target becomes the MD Gibbs residual, with the UMA-relaxed structure kept as input.
Because the MD labels provide $G$ and the equilibrated cell volume but no higher derivatives, the loss reduces to $\mathcal{L}_G$ matching the MD Gibbs residual, $\mathcal{L}_V$ matching the MD equilibrium volume ($\lambda_V = 0.05$), and a heat-capacity regularizer ($\lambda_{C_P}^\mathrm{reg} = 0.005$),
\begin{equation}
\label{eq:methods-loss-mid}
    \mathcal{L}_\mathrm{mid} = \mathcal{L}_G + \lambda_V\,\mathcal{L}_V + \lambda_{C_P}^\mathrm{reg}\,\mathcal{L}^\mathrm{reg}_{C_P}.
\end{equation}
The regularizer $\mathcal{L}^\mathrm{reg}_{C_P}$ is a smooth-$L_1$ penalty on the deviation of the predicted per-atom heat capacity from the frozen pre-trained value, with the tolerance set to $1$~J~K$^{-1}$~(mol\,atoms)$^{-1}$.
It ties the temperature dependence of the entropy and heat capacity to the well-behaved pre-trained model, while permitting an absolute anharmonic correction.
Optimization uses Adam at learning rate $2\times10^{-4}$, and the training runs for approximately ${\sim}87{,}000$ steps.

\subsection{Experimental alignment for ordered phases}

\textit{NIST--JANAF heat-capacity and entropy benchmark.}
The heat-capacity and entropy accuracy of~\cref{fig:ordered}a is measured against the NIST--JANAF thermochemical tables~\citep{chase1998janaf}, after restriction to solid phases.
The noble gases and diatomic gases, whose JANAF records are gaseous element reference states rather than solids, are excluded.
For each remaining species, we use data from $100$~K to the first melting or vaporization transition, hard-capped at $4000$~K.
On the resulting set of $255$ solids, we report the per-material mean absolute error between the recorded and predicted $C_P$ and $S$.

\textit{Holland--Powell calibration.}
For ordered phases, the post-training stage calibrates against the \texttt{ds62} internally consistent thermodynamic assessment~\citep{holland2011improved} using a predefined set of Si--Al--Mg--Ca--O minerals (the training/held-out phase split and the reaction set are tabulated in \siref{D}).
The absolute Gibbs free energy inherits a PBE-level offset from the QHA and MD stages, so the objective matches only relative stabilities and response derivatives rather than absolute $G$.
It is a smooth-$L_1$ objective combining phase-equilibrium, single-phase anchor, and reference-curve terms,
\begin{multline}
\label{eq:methods-loss-post}
    \mathcal{L}_\mathrm{post} = \mathcal{L}_\mathrm{rxn} + \lambda_\mathrm{poly}\mathcal{L}_\mathrm{poly} + \lambda_{B_0}\mathcal{L}_{B_0} + \lambda_{S_0}\mathcal{L}_{S_0} + \lambda_{V_0}\mathcal{L}_{V_0} \\
    + \lambda_{V}\mathcal{L}_{V} + \lambda_{C_P}\mathcal{L}_{C_P} + \lambda_\mathrm{shape}\mathcal{L}_\mathrm{shape} + \lambda_\mathrm{reg}\mathcal{L}_\mathrm{reg}.
\end{multline}
The dominant terms $\mathcal{L}_\mathrm{rxn}$ and $\mathcal{L}_\mathrm{poly}$ match the experimentally assessed reaction and polymorph Gibbs differences $\Delta G(T,P)$ across the preset.
The anchors $\mathcal{L}_{B_0}$, $\mathcal{L}_{S_0}$, and $\mathcal{L}_{V_0}$ match the \texttt{ds62} single-phase bulk modulus, standard entropy, and molar volume at $298.15$~K and $1$~bar through the differentiated responses of~\cref{eq:methods-S-V,eq:methods-BT}.
On the training phases, $\mathcal{L}_{V}$ and $\mathcal{L}_{C_P}$ match the \texttt{ds62} reference $V(T,P)$ and $C_P(T)$ curves, and $\mathcal{L}_\mathrm{shape}$ matches the phase-centered Gibbs surface $\hat{G}(T,P) - \hat{G}(T_0,P_0)$.
The final term $\mathcal{L}_\mathrm{reg}$ is an $L_2$ penalty toward the pre-calibration weights.
The residual bulk-modulus error against experiment is a further PBE bias shared by both earlier stages.
This stage therefore fine-tunes the full thermodynamic adapter (the encoder-attached heads together with the equation-of-state parameters), allowing the equation of state parameters to adjust.
The per-term weights are listed in \siref{D}.
Optimization uses Adam (learning rate $1\times10^{-4}$, gradient clipping at norm $5$) for $350$ steps.

\textit{\ce{SiO2} pressure--temperature phase diagram.}
The diagram (\cref{fig:ordered}j--l) is constructed from the calibrated surface by evaluating $\hat{G}(T,P)$ for the five polymorphs on a grid in temperature ($300$--$2400$~K) and pressure ($0$--$14$~GPa) and taking the minimum-$G$ phase at each grid point.
Zero contours of pairwise Gibbs free energy differences define the phase boundaries.

\subsection{Disordered solid solutions}

For a disordered solid solution, we evaluate ordered branches across composition and calibrate them against a reference: bottom-up by fine-tuning on a higher-resolution computation, or top-down by iteratively reweighting sampled configurations onto a measured phase boundary.
Both freeze the encoder backbone and update only a lightweight adapter: the bottom-up fit trains the LoRA layers, and the top-down calibration additionally updates the analytic thermodynamic heads.

\textit{Dilute solubility (bottom-up).}
The dilute solubility of \ce{Sc} in fcc \ce{Al} is determined by the branch free energies of pure \ce{Al}, the dilute \ce{Sc}-substituted solid solution, and the \ce{Al3Sc}~(L1$_2$) compound.
The per-\ce{Sc} free energy of dissolving one \ce{Al3Sc} unit into a $256$-site fcc \ce{Al} is
\begin{equation}
\label{eq:methods-alsc-diss}
    \Delta G_\mathrm{sol} = G(\ce{Al255Sc}) - G(\ce{Al3Sc}) - 252\,G(\ce{Al}),
\end{equation}
with $G(\ce{Al})$ the per-atom Gibbs free energy of pure fcc \ce{Al}, and the dilute limit follows from the ideal-dilute equilibrium
\begin{equation}
\label{eq:methods-alsc-solub}
    \frac{x_\mathrm{Sc}}{1-x_\mathrm{Sc}} = \exp\!\left(-\frac{\Delta G_\mathrm{sol}}{k_\mathrm{B}T}\right).
\end{equation}
The bottom-up calibration fine-tunes the LoRA layers on a dedicated \ce{Al}--\ce{Sc} quasi-harmonic dataset so this dilute solubility approaches the quasi-harmonic UMA reference; the vibrational entropy of dissolution, which the $0$~K static estimate omits, sets the solubility's order of magnitude~\citep{van2002effect,ozolins2001large}.
This dataset densely samples the fcc \ce{Al}--\ce{Sc} composition axis, a chemically disordered region underrepresented in pre-training.
The dataset contains $491$ substitutional supercells of up to $64$ atoms spanning the full range from pure \ce{Al} to pure \ce{Sc} across $65$ distinct \ce{Sc} fractions.
The orderings at each composition are selected for pair-correlation diversity.
Each structure carries a UMA quasi-harmonic Gibbs free energy computed by the low-fidelity procedure (\siref{B}).
Starting from the pre-trained checkpoint, the rank-$8$, scaling-$16$ adapter matches these labels with the Gibbs, volume, entropy, and heat-capacity terms of the pre-training loss~\cref{eq:methods-loss-low} ($\lambda_V = 0.1$, $\lambda_S = 0.005$, $\lambda_{C_P} = 0.001$), optimized with Adam (learning rate $1\times10^{-4}$, batch size $16$, gradient clipping at norm $10$) for $100$ epochs.

\textit{Semigrand canonical Monte Carlo.}
Equilibrium compositions within an ordered branch are sampled by semigrand canonical Monte Carlo (SGCMC)~\citep{sadigh2012scalable} on a fixed parent lattice, accelerated by well-tempered metadynamics~\citep{barducci2008well}.
At fixed species assignment $\bm{z}$, the atomic positions and cell are relaxed onto the branch representative, so vibrational and cell free energy contributions enter only through the surrogate $G_\theta$, while identity-swap moves $z_i\to z'$ are accepted with probability
\begin{equation}
\label{eq:methods-pacc}
    P_\mathrm{acc}(z_i\to z') = \min\!\big(1,\,e^{-\beta[\Delta G_\theta + \mu_{z_i}-\mu_{z'} + \Delta V_\mathrm{bias}]}\big),
\end{equation}
where $\Delta G_\theta$ is the change in the surrogate branch-conditioned Gibbs free energy of the swap and $\Delta V_\mathrm{bias}$ is the accompanying change in the metadynamics bias.
Full geometry relaxation is performed at every $10$ steps when a configuration is stored as a training or reweighting sample (after a $40\%$ burn-in).
Intermediate steps score each swap with a single surrogate evaluation on the running configuration, which enhances the sampling efficiency.
Each simulation runs $20{,}000$ steps.
Each step attempts a composition-changing identity move or a composition-conserving swap with equal probability.
Metadynamics deposits a bias $V_\mathrm{bias}$ on the composition collective variable $\bm{c}(\bm{z})$ ($0.05$~eV Gaussians every $25$ steps, bias factor $\gamma=10$); in the long-time limit $V_\mathrm{bias}\to -(1-1/\gamma)\,G_\mathrm{SGC}(\bm{c})+\mathrm{const}$, so the equilibrium concentration distribution
\begin{equation}
\label{eq:methods-pisgc}
    \pi_\mathrm{SGC}(\bm{c}) \propto \exp\!\big[-\beta\,G_\mathrm{SGC}(\bm{c};\,\bm{x}^\circ,T,P,\bm{\mu})\big]
\end{equation}
is recovered by reweighting~\citep{tiwary2015time}, and sweeping the chemical potentials $\bm{\mu}$ traces the branch-conditioned phase boundary in composition space.
The parent-lattice assumptions are detailed in \siref{B}.

\textit{Differentiable reweighting.}
The top-down calibration matches a measured phase boundary, the \ce{Au}--\ce{Pt} solvus~\citep{darling1952aupt,munster1960aupt}, by differentiable reweighting~\citep{thaler2021learning} of the stored SGCMC and metadynamics samples.
Differentiable reweighting re-scores the recorded configurations under updated adapter parameters without re-running the sampler.
At each state point, the coexistence chemical potential is determined by equating the semigrand free energies of the two basins (the common-tangent condition).
The adapter parameters then minimize a weighted error between the predicted and experimental binodal concentrations, regularized by an energy-drift penalty that keeps the importance weights valid and an $L_2$ penalty toward the pre-calibration weights.
The binodal concentrations depend on the parameters through the reweighted samples and through the implicitly differentiated coexistence root, which makes this regression loss differentiable in the adapter parameters, and optimization uses Adam (learning rate $5\times10^{-6}$, gradient clipping at norm $5$) for $160$ gradient steps per cycle.
We repeat the calibration cycle five times: within each cycle the stored samples are reused by reweighting, and between cycles the reference simulation is re-run at the updated parameters, warm-starting its metadynamics bias from the previous cycle to speed convergence.
The regression loss, the reweighting estimator, the coexistence root, and its implicit gradient are given in \siref{B}.


\section*{Code Availability}
The dataset and code to reproduce the results will be made publicly available upon acceptance of this manuscript.

\begin{acknowledgments}
The authors thank X. Fu for the discussions that inspired the conception of this project, and M. Schebek, P. Holderrieth, Y. Du, M. Cheng, K. Sheriff, Z. W. Ulissi, M. Gao, C. L. Zitnick, B. M. Wood, and others from the FAIR Chemistry team for helpful scientific disscussions.
J.N. acknowledges support from the Mathworks Fellowship.
B.D. would like to acknowledge the funding support by Shell Inc.
X.D. acknowledges funding from Amazon as part of the MIT Climate and Sustainability Consortium (MCSC).
This research used resources of the National Energy Research Scientific Computing Center (NERSC), a Department of Energy Office of Science User Facility using NERSC award ALCC-ERCAP-m5068.
\end{acknowledgments}


\section*{Competing Interests}
The authors declare no competing interests.

\bibliography{main}




\end{document}


\title{Supplementary Information for \\ Universal Thermodynamic Interatomic Potentials for Crystalline Materials}

\author{Juno Nam}
\affiliation{Department of Materials Science and Engineering, Massachusetts Institute of Technology, Cambridge, MA 02139, USA}
\author{Bowen Deng}
\affiliation{Department of Materials Science and Engineering, Massachusetts Institute of Technology, Cambridge, MA 02139, USA}
\author{Xiaochen Du}
\affiliation{Department of Materials Science and Engineering, Massachusetts Institute of Technology, Cambridge, MA 02139, USA}
\author{Luis Barroso-Luque}
\affiliation{Fundamental AI Research, Meta, San Francisco, CA 94105, USA}
\author{Benjamin Kurt Miller}
\email{bkmi@meta.com}
\affiliation{Fundamental AI Research, Meta, San Francisco, CA 94105, USA}
\author{Rafael G\'omez-Bombarelli}
\email{rafagb@mit.edu}
\affiliation{Department of Materials Science and Engineering, Massachusetts Institute of Technology, Cambridge, MA 02139, USA}

\date{\today}

\maketitle
{
\setstretch{1.25}
\onecolumngrid
\begingroup
\hypersetup{linkcolor=black}
\tableofcontents
\endgroup
\clearpage
\twocolumngrid
}


\section{Theoretical Formalism and Statistical Basis}
\label{sec:formalism}

\begin{table}[!ht]
\caption{
\textbf{Summary of the notation.}
}
\label{table:notation}
\begin{tblr}{colspec=cX,rowsep=0.5pt,colsep=6pt,vline{2}}
\toprule
Symbol & Description \\
\midrule
$N$ & Number of atoms in a crystal \\
$\bm{x}$ & Crystal structure $(\bm{L}, \bm{z}, [\bm{r}])$ \\
$\bm{x}^\circ$ & Branch representative (relaxed structure) \\
$\bm{L}$ & Cell matrix ($\mathrm{GL}^+(3, \mathbb{R})$) \\
$V$ & Cell volume ($\det \bm{L} > 0$) \\
$\bm{z}$ & Atomic species in a crystal ($\mathcal{A}^N$) \\
$\bm{m}$ & Atomic masses ($\mathbb{R}_{>0}^N$) \\
$\bm{r}$ & Cartesian atomic positions ($\mathbb{R}^{3 \times N}$) \\
$\bm{c}$ & Composition on a parent lattice \\
\midrule
$O^\circ$ & Relaxed/reference observable $O$ \\
$O^\ast$ & Equilibrium (ensemble) average of $O$ \\
$\hat{O}$ & Predicted observable $O$ \\
\midrule
$P$ & External pressure \\
$T$ & External temperature \\
$\beta$ & Inverse temperature $1 / k_\mathrm{B} T$ \\
$U$ & Potential energy (of a configuration) \\
$U^\circ$ & Static reference energy $U(\bm{x}^\circ)$ \\
$F$ & Helmholtz free energy ($NVT$) \\
$G$ & Gibbs free energy ($NPT$) \\
$G^\delta$ & Thermal residual ($G - U^\circ$) \\
\midrule
$C_P$ & Isobaric heat capacity \\
$V_0$ & Equilibrium volume at $P = 0$ \\
$B_0$ & Isothermal bulk modulus at $P = 0$ \\
$\alpha$ & Volumetric thermal expansion coefficient \\
$S(298~\mathrm{K})$ & Standard entropy at $298$~K \\
\midrule
$\lambda$ & Switching (coupling) coordinate ($[0, 1]$) \\
$\gamma$ & Well-tempered metadynamics bias factor \\
$\mu_z$ & Chemical potential of species $z$ \\
$\Delta\mu$ & Semigrand chemical potential difference \\
$\theta$ & TIP parameters \\
\bottomrule
\end{tblr}
\end{table}

\subsection{Notation}

\cref{table:notation} summarizes the notation used throughout this work.
A periodic crystal of $N$ atoms is represented by the triple
\begin{equation}
    \bm{x} = (\bm{L}, \bm{z}, [\bm{r}]) \in \mathcal{X}_N,
\end{equation}
where $\bm{L} \in \mathrm{GL}^+(3, \mathbb{R})$ is the cell matrix whose columns are the lattice vectors.
The species $\bm{z} = (z_1, \dots, z_N) \in \mathcal{A}^N$ are drawn from the periodic table $\mathcal{A}$ (a finite set of chemical elements).
The symbol $[\bm{r}]$ denotes the periodic Cartesian configuration of the $N$ atoms, i.e., the equivalence class of $\bm{r} \in \mathbb{R}^{3 \times N}$ modulo lattice translations.
We fix the residual $\mathrm{O}(3)$ gauge associated with global rotations and reflections by restricting $\bm{L}$ to upper-triangular form with positive diagonal entries.

The cell matrix induces the Bravais lattice $\bm{L}\mathbb{Z}^3$ with fundamental-cell volume $V := \det \bm{L} > 0$, and each periodic position $[\bm{r}_i]$ has a unique representative in that domain.

For a structure-dependent observable $O(\bm{x})$, we write $O^\circ := O(\bm{x}^\circ)$ for its zero-temperature value at the branch representative, and
\begin{equation}
    O^\ast := \mathbb{E}_{\bm{x} \sim \pi_{NPT}}\!\left[ O(\bm{x}) \right]
\end{equation}
for its expectation under the branch-conditioned $NPT$ Gibbs measure $\pi_{NPT}(\bm{x};\, \bm{x}^\circ, T, P) \propto e^{-\beta\,(U(\bm{x}) + PV)}$ supported on $\mathcal{B}(\bm{x}^\circ)$ (the branch, defined in the Assumptions below), formally normalized in~\cref{eq:Delta-NPT}.

\subsection{Assumptions}

The thermodynamic interatomic potential (TIP) models the Gibbs free energy of an ordered bulk crystal branch indexed by the reference structure $\bm{x}^\circ$.
To make this precise, let
\begin{equation}
    \mathcal{X}_N^\circ \subset \mathcal{X}_N
\end{equation}
denote the subset of reference (relaxed) configurations.
We then introduce a deterministic, idempotent \emph{branch map}
\begin{equation}
    \mathfrak{R}_\mathrm{sym}: \mathcal{X}_N \to \mathcal{X}_N^\circ, \qquad \mathfrak{R}_\mathrm{sym} \circ \mathfrak{R}_\mathrm{sym} = \mathfrak{R}_\mathrm{sym},
\end{equation}
that maps each periodic structure $\bm{x} \in \mathcal{X}_N$ to its branch representative $\mathfrak{R}_\mathrm{sym}(\bm{x}) \in \mathcal{X}_N^\circ$.
In practice, $\mathfrak{R}_\mathrm{sym}$ is implemented by minimizing the base potential energy $U$ (the UMA potential for TIP[UMA]) while constraining the reference space group, so that $\mathfrak{R}_\mathrm{sym}(\bm{x})$ is a stationary point
\begin{equation}
    \nabla_{\bm{r}, \bm{L}} U\big|_{\mathfrak{R}_\mathrm{sym}(\bm{x})} \approx 0
\end{equation}
under the imposed constraint.
For ordinary phases, the constraint is inactive and $\mathfrak{R}_\mathrm{sym}$ coincides with unconstrained 0\,K relaxation.
For dynamically stabilized phases such as bcc Ti above the martensitic transition, the constraint is essential because the static high-symmetry archetype is not itself a mechanical minimum.
The preimage
\begin{equation}
    \mathcal{B}(\bm{x}^\circ) := \mathfrak{R}_\mathrm{sym}^{-1}(\bm{x}^\circ) = \{\bm{x} \in \mathcal{X}_N: \mathfrak{R}_\mathrm{sym}(\bm{x}) = \bm{x}^\circ \}
\end{equation}
is the \emph{branch} (or basin) associated with $\bm{x}^\circ$, and the family $\{ \mathcal{B}(\bm{x}^\circ) \}_{\bm{x}^\circ \in \mathcal{X}_N^\circ}$ partitions $\mathcal{X}_N$ by construction.

The formulation uses three assumptions:
\begin{enumerate}[label=(A\arabic*),leftmargin=*,align=left]
\item \textit{Branch representability.}
The reference structure $\bm{x}^\circ \in \mathcal{X}_N^\circ$ is a fixed point of $\mathfrak{R}_\mathrm{sym}$, i.e., $\mathfrak{R}_\mathrm{sym}(\bm{x}^\circ) = \bm{x}^\circ$, and a stationary point of $U$ under the symmetry constraint defining $\mathfrak{R}_\mathrm{sym}$, but it need not be an unconstrained local minimum.

\item \textit{Finite-$(T, P)$ branch stability.}
At the queried $(T, P)$, the branch $\mathcal{B}(\bm{x}^\circ)$ is a stable or metastable $NPT$ basin: the system equilibrates within $\mathcal{B}(\bm{x}^\circ)$ on the simulation timescale, while escape to other branches is exponentially rare over that timescale.

\item \textit{Fixed-branch scope.}
A TIP models the \emph{branch-conditioned} Gibbs free energy $G(\bm{x}^\circ; T, P)$ obtained by restricting the $NPT$ partition function to $\bm{x} \in \mathcal{B}(\bm{x}^\circ)$.
Competing branches are compared after evaluating $G$ for their respective representatives.
\end{enumerate}

A1 treats $\bm{x}^\circ$ as a branch label rather than requiring an unconstrained mechanical minimum, admitting dynamically stabilized phases.
A2 makes the conditioned ensemble well-defined, so that long-time molecular dynamics (MD) initialized in $\mathcal{B}(\bm{x}^\circ)$ samples the branch ergodically.
A3 limits the target to a single branch, and therefore requires an unambiguous symmetry constraint for $\mathfrak{R}_\mathrm{sym}$.

\subsection{Branch-Conditioned Gibbs Free Energy}
\label{sec:branch-G}

Following the fixed-ordering coarse-graining framework of vibrational alloy thermodynamics~\citep{van2002effect}, a TIP targets the Gibbs free energy of a crystal conditioned on its branch label $\bm{x}^\circ \in \mathcal{X}_N^\circ$, rather than the free energy summed over all branches.

At fixed cell $\bm{L}$, the branch-conditioned canonical ($NVT$) partition function is
\begin{equation}
\label{eq:Z-NVT}
    Z(\bm{x}^\circ, \bm{L};\, T) := \frac{1}{\sigma(\bm{z})\,\Lambda_{\bm{m}}^{3N}} \int_{\mathcal{B}_{\bm{L}}(\bm{x}^\circ)} \exp\!\big[ -\beta U(\bm{x}) \big]\, \mathrm{d}\bm{r},
\end{equation}
where $\beta := 1/k_\mathrm{B} T$, $\mathcal{B}_{\bm{L}}(\bm{x}^\circ) := \mathcal{B}(\bm{x}^\circ) \cap \{\bm{x}\,:\,\text{cell}(\bm{x}) = \bm{L}\}$ is the slice of the branch at fixed cell, $\sigma(\bm{z})$ is the indistinguishability factor for permutations of identical species, and $\Lambda_{\bm{m}}^{3N} := \prod_{i=1}^N (h^2/2\pi m_i k_\mathrm{B} T)^{3/2}$ is the species-resolved thermal de Broglie prefactor obtained by integrating out atomic momenta.
The associated branch-conditioned Helmholtz free energy is
\begin{equation}
\label{eq:F-branch}
    F(\bm{x}^\circ, \bm{L};\, T) := -k_\mathrm{B} T \ln Z(\bm{x}^\circ, \bm{L};\, T).
\end{equation}

The branch-conditioned isothermal-isobaric ($NPT$) partition function is the Laplace transform of $Z$ over the cell degrees of freedom at fixed pressure,
\begin{equation}
\label{eq:Delta-NPT}
    \Delta(\bm{x}^\circ;\, T, P) := \int e^{-\beta P V}\, Z(\bm{x}^\circ, \bm{L};\, T)\, \mathrm{d}\bm{L},
\end{equation}
where the integral runs over cell matrices $\bm{L}$ consistent with the branch, in the upper-triangular gauge introduced in the Notation subsection, and $V = \det \bm{L}$.
Here $\mathrm{d}\bm{L}$ denotes a fixed reference measure on cell matrices.
Its choice and the discrete multiplicity of equivalent lattice bases enter $G$ only through a subextensive normalization that is negligible per atom and cancels in the branch differences and $(T, P)$-derivatives used here.
The corresponding \emph{branch-conditioned Gibbs free energy} is
\begin{equation}
\label{eq:G-branch}
    G(\bm{x}^\circ;\, T, P) := -k_\mathrm{B} T \ln \Delta(\bm{x}^\circ;\, T, P).
\end{equation}
By construction, $G(\bm{x}^\circ;\, T, P)$ is a branchwise quantity, obtained by marginalizing over the vibrational and cell-shape fluctuations within a single branch $\mathcal{B}(\bm{x}^\circ)$.
Summing over the disjoint branches gives the free energy over the full configuration space,
\begin{equation}
\label{eq:G-total}
    G(T, P) = -k_\mathrm{B} T \ln \sum_{\bm{x}^\circ \in \mathcal{X}_N^\circ} \exp\!\big[ -\beta\, G(\bm{x}^\circ;\, T, P) \big].
\end{equation}
For a chemically disordered system, this branch sum is evaluated by sampling the disjoint branches on a parent lattice, as in the main-text solid solution examples.
We carry this out by configurational sampling on a parent lattice, as for the disordered solid solutions of the main text.

Equilibrium observables within the branch are obtained as standard thermodynamic derivatives of $G(\bm{x}^\circ;\, T, P)$,
\begin{equation}
\label{eq:S-V}
    S = -\left(\frac{\partial G}{\partial T}\right)_{\!P}, \qquad V = \left(\frac{\partial G}{\partial P}\right)_{\!T}.
\end{equation}
For modeling purposes, we split the target into a static reference and a thermodynamic residual,
\begin{equation}
\label{eq:G-residual}
    G(\bm{x}^\circ;\, T, P) = U^\circ + G^\delta(\bm{x}^\circ;\, T, P),
\end{equation}
where $U^\circ := U(\bm{x}^\circ)$ is the zero-temperature potential energy of the branch representative and $G^\delta$ is the thermodynamic residual that contains all finite-$(T, P)$ contributions.

\subsection{Statistical Basis of the Learned Free Energy}

The quasi-harmonic approximation (QHA) and nonequilibrium MD simulation free energies use different nuclear statistics.
After removal of the zero-point term, the QHA label in~\cref{eq:F-vib} is the quantum thermal-occupation free energy $\sum_{\bm{q}} w_{\bm{q}} \sum_\nu k_\mathrm{B} T \ln(1 - e^{-\beta \hbar \omega_{\bm{q}\nu}})$, whereas the Frenkel--Ladd free energy in~\cref{eq:F-fl} is classical: it is built on the classical Einstein reference of~\cref{eq:F-harm} and sampled by classical molecular dynamics.
The two limits agree as $\beta \hbar \omega \to 0$, where $k_\mathrm{B} T \ln(1 - e^{-\beta \hbar \omega}) \to k_\mathrm{B} T \ln(\beta \hbar \omega)$ reproduces the classical form of~\cref{eq:F-harm}.
At low temperatures, they differ: the quantum heat capacity approaches zero, whereas the classical harmonic heat capacity approaches $3 N k_\mathrm{B}$.
The analytic head combines these targets by retaining the quantum Einstein term $3\, k_\mathrm{B} T \ln(1 - e^{-\theta_\mathrm{E} / T})$, while fitting MD-derived anharmonic corrections over the sampled temperature range.
Mid-trained TIP[UMA] therefore retains nuclear quantum effects (except the zero-point offset) at the harmonic level and incorporates classical anharmonic corrections from MD.

\section{Derivations of the Free Energy and Sampling Methods}

\subsection{Quasi-Harmonic Approximation}
\label{sec:qha}

The QHA~\citep{baroni2001phonons,togo2015first} provides the low-fidelity approximation for $G(\bm{x}^\circ;\, T, P)$ that retains volume-dependent vibrational physics while neglecting anharmonic mode coupling.
We use it as the low-fidelity training signal for TIP[UMA].

For a fixed cell $\bm{L}$, we expand the potential energy around the cell-constrained relaxed positions $\bm{r}^\circ(\bm{L}) := \arg\min_{\bm{r}} U(\bm{r};\, \bm{L})$ to second order,
\begin{equation}
\label{eq:U-harmonic-qha}
    U(\bm{r};\, \bm{L}) \approx U^\circ(\bm{L}) + \frac{1}{2}\,\Delta\bm{r}^\top \Phi(\bm{L})\, \Delta\bm{r},
\end{equation}
where $\Delta\bm{r} := \bm{r} - \bm{r}^\circ(\bm{L})$, $U^\circ(\bm{L}) := U(\bm{r}^\circ(\bm{L});\, \bm{L})$ is the cell-constrained static energy, and $\Phi(\bm{L})$ is the interatomic force-constant matrix at $\bm{r}^\circ(\bm{L})$.
By lattice periodicity the mass-weighted force constants block-diagonalize over the Brillouin zone: at each wavevector $\bm{q}$ the dynamical matrix $\widetilde{\Phi}(\bm{q};\, \bm{L}) \in \mathbb{C}^{3n \times 3n}$ ($n$ atoms per unit cell) has eigenvalues $\omega_{\bm{q}\nu}^2(\bm{L})$ indexed by the branch $\nu = 1, \dots, 3n$.

Within this harmonic ansatz at fixed cell, the canonical partition function factorizes into independent quantum harmonic oscillators, giving the closed-form Helmholtz free energy
\begin{equation}
\label{eq:F-qha}
    F_\mathrm{QHA}(\bm{L}, T) = U^\circ(\bm{L}) + F_\mathrm{vib}(\bm{L}, T),
\end{equation}
with the vibrational contribution
\begin{multline}
\label{eq:F-vib}
    F_\mathrm{vib}(\bm{L}, T) = \sum_{\bm{q}} w_{\bm{q}} \sum_{\nu=1}^{3n} \bigg[ \frac{\hbar\, \omega_{\bm{q}\nu}(\bm{L})}{2} \\
    + k_\mathrm{B} T \ln\!\Big(1 - e^{-\beta \hbar \omega_{\bm{q}\nu}(\bm{L})}\Big) \bigg],
\end{multline}
per unit cell, where the sum runs over a regular mesh of wavevectors $\bm{q}$ in the first Brillouin zone with weights normalized to $\sum_{\bm{q}} w_{\bm{q}} = 1$.
The first term is the zero-point energy and the second is the thermal occupation contribution.
The cell dependence of $\omega_{\bm{q}\nu}(\bm{L})$ accounts for the thermal expansion and Gr\"uneisen-type effects that a fixed-cell harmonic approximation omits.
The force constants are obtained from finite displacements in the relaxed supercell and Fourier-interpolated onto the Brillouin-zone mesh with Phonopy~\citep{togo2015first}.
We discard modes below a small near-zero cutoff, including the three acoustic $\Gamma$-point modes and small imaginary modes caused by numerical noise, to avoid singular occupation terms.

The branch-conditioned QHA Gibbs free energy is obtained by minimizing $F(L, T) + PV$ over cells consistent with the branch,
\begin{equation}
\label{eq:G-qha}
    G_\mathrm{QHA}(\bm{x}^\circ;\, T, P) := \min_{\bm{L}} \big[ F_\mathrm{QHA}(\bm{L}, T) + P V \big],
\end{equation}
where the minimization runs over cells consistent with the branch $\mathcal{B}(\bm{x}^\circ)$.
In practice we restrict the search to an isotropic-strain family $\bm{L}(\eta) := (1 + \eta)\, \bm{L}^\circ$ around the branch representative, evaluate $F_\mathrm{QHA}$ on a strain grid, and fit the resulting $F_\mathrm{QHA}(V, T)$ with a Vinet equation of state~\citep{vinet1987temperature} to perform the minimization analytically.
Subtracting the static reference yields the QHA approximation to the thermodynamic residual,
\begin{equation}
\label{eq:Gdelta-qha}
    G^\delta_\mathrm{QHA}(\bm{x}^\circ;\, T, P) := G_\mathrm{QHA}(\bm{x}^\circ;\, T, P) - U^\circ,
\end{equation}
which is the regression target for the low-fidelity training signal.

Within the isotropic-strain restriction, QHA is exact for a harmonic vibrational Hamiltonian and captures leading volumetric anharmonicity through $\omega_{\bm{q}\nu}(\bm{L})$.
It neglects anisotropic cell relaxation and mode-mode coupling, and breaks down when any $\omega_{\bm{q}\nu}^2(\bm{L}) < 0$ along the relevant cell trajectory~\citep{van2015free}.
For dynamically stabilized branches covered by assumption (A1), the static reference $\bm{x}^\circ$ is itself unstable under unconstrained relaxation, and QHA must be augmented or replaced by a finite-temperature method such as Frenkel--Ladd switching (\cref{sec:frenkel-ladd}).

\subsection{Frenkel--Ladd Switching}
\label{sec:frenkel-ladd}

Frenkel--Ladd switching~\citep{frenkel1984new} computes the absolute Helmholtz free energy of a single crystalline branch by thermodynamic integration along a path that connects the physical Hamiltonian to an analytically tractable harmonic reference.
We follow the optimized switching variant of~\citet{de1999optimized}.

Let $U(\bm{r})$ denote the physical potential energy of the branch evaluated at the cell $\bm{L}^\ast$ equilibrated at the target $(T, P)$ under the branch-conditioned $NPT$ ensemble, and define the Einstein crystal reference
\begin{equation}
\label{eq:U-harm}
    U_\mathrm{harm}(\bm{r}) := \frac{1}{2} \sum_{i=1}^N k_i\, \big\| \bm{r}_i - \bm{r}_i^\circ \big\|^2,
\end{equation}
where $\bm{r}_i^\circ$ are the relaxed atomic positions of $\bm{x}^\circ$ and $k_i$ are species-resolved spring constants.
The mixed Hamiltonian
\begin{equation}
\label{eq:H-fl}
    U(\bm{r}; \lambda) := (1 - \lambda)\, U(\bm{r}) + \lambda\, U_\mathrm{harm}(\bm{r}), \qquad \lambda \in [0, 1],
\end{equation}
interpolates between the two endpoints.
Letting $\pi_\lambda(\bm{r}) \propto e^{-\beta U(\bm{r}; \lambda)}$ denote the canonical ($NVT$) measure under the mixed Hamiltonian at fixed $\lambda$, the reversible-work theorem reads
\begin{equation}
\label{eq:DeltaF-rev}
    F_\mathrm{harm} - F = \int_0^1 \mathbb{E}_{\bm{r} \sim \pi_\lambda}\!\left[ U_\mathrm{harm}(\bm{r}) - U(\bm{r}) \right] \mathrm{d}\lambda.
\end{equation}
For a finite-rate switching trajectory in which $\lambda(t)$ is driven from $0$ to $1$ over a time $\tau_s$, the accumulated nonequilibrium work
\begin{equation}
\label{eq:W-fl}
    W^s := \int_0^{\tau_s}\!\dot{\lambda}(t)\, \big[ U_\mathrm{harm}(\bm{r}(t)) - U(\bm{r}(t)) \big]\, \mathrm{d}t
\end{equation}
is a path-dependent random variable whose distribution depends on the switching protocol; let $\mathbb{P}_{0 \to 1}$ and $\mathbb{P}_{1 \to 0}$ denote the forward and backward switching path measures, each initialized from the corresponding equilibrium endpoint.
The driving follows a smooth polynomial schedule~\citep{deKoning1996einstein,nam2025alchemical} in the reduced time $\tilde{t} = t/\tau_s$,
\begin{equation}
\label{eq:lambda-schedule}
    \lambda(\tilde{t}) = \tilde{t}^5\big(70\tilde{t}^4 - 315\tilde{t}^3 + 540\tilde{t}^2 - 420\tilde{t} + 126\big),
\end{equation}
whose slope $\mathrm{d}\lambda/\mathrm{d}\tilde{t} = 630\,\tilde{t}^4(1-\tilde{t})^4$ vanishes at both endpoints and reduces the dissipation relative to a linear ramp.
The Jarzynski--Crooks dissipation inequality $\mathbb{E}_{\mathbb{P}_{0 \to 1}}[W^s] \ge F_\mathrm{harm} - F$ holds with equality only in the quasistatic limit.
Forward and backward trajectories are combined symmetrically~\citep{de1999optimized} so that the leading-order dissipation cancels, giving the bidirectional estimator
\begin{equation}
\label{eq:DeltaF-sym}
    \Delta F = \frac{1}{2}\big( \mathbb{E}_{\mathbb{P}_{0 \to 1}}[W^s] - \mathbb{E}_{\mathbb{P}_{1 \to 0}}[W^s] \big)
\end{equation}
and the bidirectional dissipation diagnostic
\begin{equation}
\label{eq:Ed}
    E_\mathrm{d} := \frac{1}{2}\big( \mathbb{E}_{\mathbb{P}_{0 \to 1}}[W^s] + \mathbb{E}_{\mathbb{P}_{1 \to 0}}[W^s] \big).
\end{equation}

The reference free energy is known in closed form: the classical Einstein crystal of $N$ independent three-dimensional oscillators with frequencies $\omega_i := \sqrt{k_i / m_i}$ has
\begin{equation}
\label{eq:F-harm}
    F_\mathrm{harm}(T) = 3\, k_\mathrm{B} T \sum_{i=1}^N \ln\!\left( \frac{\hbar \omega_i}{k_\mathrm{B} T} \right),
\end{equation}
so the absolute Helmholtz free energy of the physical branch follows from~\cref{eq:DeltaF-sym} as
\begin{equation}
\label{eq:F-fl}
    F(\bm{x}^\circ, \bm{L}^\ast;\, T) = F_\mathrm{harm}(T) - \Delta F.
\end{equation}
Pinning each atom to its lattice site $\bm{r}_i^\circ$ in $U_\mathrm{harm}$ also fixes the center of mass of the reference, a constraint absent from the unconstrained crystal partition function.
The corresponding finite-size correction~\citep{frenkel1984new} is
\begin{multline}
\label{eq:F-com}
    F_\mathrm{COM}(T) = -k_\mathrm{B} T \ln\!\bigg[ V_\mathrm{prim} \bigg( 2\pi k_\mathrm{B} T \sum_{i=1}^N \frac{\nu_i^2}{k_i} \bigg)^{\!-3/2} \bigg],
\end{multline}
where $V_\mathrm{prim}$ is the primitive-cell volume and $\nu_i := m_i / \sum_j m_j$ are the mass fractions.
Combining the absolute Helmholtz free energy, the $G = F + P V$ conversion at the equilibrated cell, and the COM correction yields the Frenkel--Ladd estimate of the branch-conditioned Gibbs free energy
\begin{equation}
\label{eq:G-fl}
    G_\mathrm{FL}(\bm{x}^\circ;\, T, P) = F_\mathrm{harm}(T) - \Delta F + P V + F_\mathrm{COM}(T).
\end{equation}
Because this $G = F + P V$ conversion evaluates the Helmholtz free energy at the single equilibrated cell $\bm{L}^\ast$ rather than integrating over cell fluctuations, the estimate~\cref{eq:G-fl} is an approximation to the full branch-conditioned $NPT$ Gibbs free energy~\cref{eq:G-branch}.
This is the same cell-fluctuation approximation adopted in the CALPHY workflow~\citep{menon2021automated}.
The neglected cell-fluctuation contribution arises from the few cell degrees of freedom, so it is of order $k_\mathrm{B} T$ per simulation cell and subextensive, hence negligible per atom for the ${\sim}1{,}000$-atom supercells used here.
It can nonetheless grow for strongly flexible or anharmonic cells.

\subsection{Alchemical Switching}
\label{sec:alchemical}

Alchemical switching estimates the Gibbs free energy difference between two potential energy surfaces $U^A$ and $U^B$ at the same thermodynamic state $(\bm{x}^\circ;\, T, P)$, without requiring an absolute reference for either.
The mixed Hamiltonian
\begin{equation}
\label{eq:H-alc}
    U(\bm{r}; \lambda) := (1 - \lambda)\, U^A(\bm{r}) + \lambda\, U^B(\bm{r}), \qquad \lambda \in [0, 1],
\end{equation}
interpolates between the two surfaces at fixed species and temperature, holding species and temperature fixed while allowing the cell variables to respond to $(T, P)$.
Both legs are run in the $NPT$ ensemble.
The same forward-backward symmetrization of~\cref{eq:DeltaF-sym}, with switching path measures $\mathbb{P}_{A \to B}$ and $\mathbb{P}_{B \to A}$ replacing $\mathbb{P}_{0 \to 1}$ and $\mathbb{P}_{1 \to 0}$, now yields the Gibbs free energy difference
\begin{equation}
\label{eq:DeltaG-alc}
\begin{aligned}
    \Delta G^{A \to B}(\bm{x}^\circ;\, T, P) &:= G^B - G^A \\
    &= \frac{1}{2}\big( \mathbb{E}_{\mathbb{P}_{A \to B}}[W^s] - \mathbb{E}_{\mathbb{P}_{B \to A}}[W^s] \big),
\end{aligned}
\end{equation}
where each switching work is defined as in~\cref{eq:W-fl} with the integrand $U_\mathrm{harm} - U$ replaced by $U^B - U^A$.

Combining alchemical switching with an absolute Frenkel--Ladd estimate~\cref{eq:G-fl} performed against a faster potential $A$ transfers the absolute Gibbs free energy onto a target surface $B$ via the triangle relation
\begin{equation}
\label{eq:G-transfer}
    G^B(\bm{x}^\circ;\, T, P) = G^A(\bm{x}^\circ;\, T, P) + \Delta G^{A \to B}(\bm{x}^\circ;\, T, P).
\end{equation}
We use this with $A =$ Orb-v3 (the faster reference) and $B =$ UMA (the target): an absolute Frenkel--Ladd reference is computed on a faster potential, and a single alchemical step then transfers it onto the target surface, sidestepping the computational cost of a direct Frenkel--Ladd calculation on the latter.
Because Orb-v3 and UMA are trained on overlapping density functional theory data, their potential energy surfaces are closely matched, so the dissipation is much smaller than that of the Frenkel--Ladd switching.

\subsection{Semigrand Canonical Monte Carlo}
\label{sec:sgcmc}

We combine semigrand canonical Monte Carlo (SGCMC)~\citep{sadigh2012scalable} with well-tempered metadynamics (WTMetaD)~\citep{barducci2008well} to sample equilibrium species concentrations on a fixed parent lattice.
The parent lattice is a structural template defined by $\bm{L}^\circ$ and $N$ atomic sites $\bm{r}^\circ = (\bm{r}_1^\circ, \dots, \bm{r}_N^\circ)$ shared by all species assignments under consideration.
For each $\bm{z} \in \mathcal{A}^N$, the corresponding branch representative is
\begin{equation}
\label{eq:xz-rep}
    \bm{x}^\circ_{\bm{z}} := \mathfrak{R}_\mathrm{sym}\!\big( (\bm{L}^\circ, \bm{z}, [\bm{r}^\circ]) \big) \in \mathcal{X}_N^\circ,
\end{equation}
and the SGCMC ensemble is supported on the parent-lattice union $\bigcup_{\bm{z} \in \mathcal{A}^N} \mathcal{B}(\bm{x}^\circ_{\bm{z}})$.
The Markov chain samples only the discrete site-occupancy variable $\bm{z}$: the continuous coordinates $(\bm{r}, \bm{L})$ are relaxed onto the branch representative and contribute only through the marginalized branch-conditioned Gibbs free energy
\begin{equation}
\label{eq:G-z}
    G(\bm{z};\, T, P) := G(\bm{x}^\circ_{\bm{z}};\, T, P)
\end{equation}
of~\cref{eq:G-branch}.
We assume the parent lattice is preserved across all assignments under consideration, i.e., $\mathfrak{R}_\mathrm{sym}$ acts site-locally without breaking the shared lattice topology.
This excludes vacancies, interstitials, and symmetry-changing reconstructions, which change the site count or lattice topology and lie outside this scope.

Within this formalism, SGCMC samples a Gibbs measure at fixed $N$, $T$, $P$, and chemical potentials $\boldsymbol{\mu} := (\mu_z)_{z \in \mathcal{A}}$, while the assignment $\bm{z} \in \mathcal{A}^N$ fluctuates; metadynamics promotes sampling across the relevant composition basins so that all phases of interest are visited within a single trajectory.

The SGC ensemble is sampled over the discrete assignment $\bm{z}$ by \emph{identity swap moves}, with the configurational coordinates $(\bm{r}, \bm{L})$ relaxed onto the branch representative: select a site $i$ uniformly at random, propose changing its species $z_i \to z' \in \mathcal{A}$, and accept with probability
\begin{equation}
\label{eq:swap-accept}
    P_\mathrm{acc}(z_i \to z') = \min\!\Big(1,\, e^{-\beta\, [\Delta G_\theta + \mu_{z_i} - \mu_{z'} + \Delta V_\mathrm{bias}]}\Big),
\end{equation}
where $\Delta G_\theta$ is the change in the surrogate branch-conditioned Gibbs free energy $G_\theta$ under the swap, evaluated on the running configuration, and $\Delta V_\mathrm{bias}$ is the accompanying change in the well-tempered metadynamics bias~\cref{eq:Vbias} on the composition collective variable.
Only chemical-potential differences are physical, so we fix a reference species $z_0 \in \mathcal{A}$ and parametrize the ensemble by $\Delta\mu_z := \mu_z - \mu_{z_0}$ for $z \ne z_0$.
Detailed balance with~\cref{eq:swap-accept} yields, at fixed bias, the biased semigrand measure $\pi_\theta(\bm{z}) \propto \exp\!\big[ -\beta\big( G_\theta(\bm{z};\, T, P) - \textstyle\sum_{z \in \mathcal{A}} \mu_z N_z(\bm{z}) \big) - \beta\, V_\mathrm{bias}(\bm{c}(\bm{z})) \big]$ as the stationary distribution of the Markov chain~\citep{sadigh2012scalable}; the unbiased semigrand distribution~\cref{eq:c-eq} is recovered by reweighting.

The acceptance ratio of~\cref{eq:swap-accept} requires only the surrogate Gibbs change $\Delta G_\theta$ and the bias change of the proposed swap, obtained from a single surrogate evaluation on the running configuration.
The full symmetry-constrained relaxation occurs only on logging steps, i.e., when the configurations are retained as training or reweighting samples.
Intermediate proposals are not relaxed, which reduces the computational cost of sampling, while all stored configurations are fully relaxed.
For intermediate proposals, evaluating $G_\theta$ on the running configuration approximates the relaxed branch free energy $G(\bm{x}^\circ_{\bm{z}})$ of~\cref{eq:G-z}.
The acceptance probability depends only on the free energy difference of two configurations that differ by a single swap, for which the relaxation contribution largely cancels, and every stored sample carries the fully relaxed value used in the reweighting of~\cref{sec:difftre}.
In the production runs, each step attempts a composition-changing identity move or a composition-conserving swap with equal probability.
The well-tempered bias of~\cref{eq:Vbias} deposits $0.05$~eV Gaussians every $25$ steps with bias factor $\gamma = 10$.
Configurations are retained for relaxation and labeling every $10$ steps after a $40\%$ burn-in.

To accelerate composition sampling, define the composition collective variable
\begin{equation}
\label{eq:cv}
    \bm{c}(\bm{z}) := \frac{1}{N}\big( N_z(\bm{z}) \big)_{z \in \mathcal{A}}, \qquad N_z(\bm{z}) := \sum_{i=1}^N \bm{1}\{ z_i = z \},
\end{equation}
which lives on the $(|\mathcal{A}| - 1)$-simplex.
Well-tempered metadynamics~\citep{barducci2008well} deposits Gaussian kernels along the CV trajectory,
\begin{equation}
\label{eq:Vbias}
    V_\mathrm{bias}(\bm{c}, t) = \sum_{t' \le t} w(t')\, \exp\!\left[ -\frac{\| \bm{c} - \bm{c}(t') \|^2}{2 \sigma^2} \right],
\end{equation}
with deposition heights rescaled by the running bias,
\begin{equation}
\label{eq:wt-height}
    w(t) = w_0\, \exp\!\left[ -\frac{V_\mathrm{bias}(\bm{c}(t),\, t)}{(\gamma - 1)\, k_\mathrm{B} T} \right].
\end{equation}
The bias factor $\gamma > 1$ sets an effective CV temperature $T_\mathrm{eff} := \gamma T$; in the long-time limit the bias converges to
\begin{equation}
\label{eq:Vbias-converged}
    V_\mathrm{bias}(\bm{c}) \xrightarrow{t \to \infty} -\bigg(1 - \frac{1}{\gamma}\bigg) G_\mathrm{SGC}(\bm{c};\, \bm{x}^\circ, T, P, \boldsymbol{\mu}) + \mathrm{const},
\end{equation}
where $G_\mathrm{SGC}(\bm{c};\, \cdots)$ is the projection of the SGC Gibbs free energy onto the composition CV at fixed branch and external conditions.

The equilibrium concentration distribution at $(T, P, \boldsymbol{\mu})$ on the branch $\mathcal{B}(\bm{x}^\circ)$ is recovered by reweighting according to~\cref{eq:Vbias-converged},
\begin{equation}
\label{eq:c-eq}
    \pi_\mathrm{SGC}(\bm{c};\, \bm{x}^\circ, T, P, \boldsymbol{\mu}) \propto \exp\!\big[ -\beta\, G_\mathrm{SGC}(\bm{c};\, \bm{x}^\circ, T, P, \boldsymbol{\mu}) \big],
\end{equation}
and is reweighted from the biased trajectory using the time-independent estimator of~\citet{tiwary2015time}.
Equilibrium concentrations are identified with the basin-conditioned mean concentrations of $\pi_\mathrm{SGC}$, and sweeping $\boldsymbol{\mu}$ at fixed $(T, P)$ traces out the branch-conditioned phase boundary in composition space.

\subsection{Differentiable Reweighting}
\label{sec:difftre}

The parameters $\theta$ of a Gibbs free energy model $G_\theta(\bm{z};\, T, P)$ are fine-tuned to reproduce experimental binodal concentrations by differentiable reweighting~\citep{thaler2021learning} of stored SGCMC plus WTMetaD samples.
For clarity, we specialize to a binary alloy with species $\mathcal{A} = \{A, B\}$, scalar concentration $c(\bm{z}) := N_B(\bm{z})/N$, and chemical potential difference $\Delta\mu := \mu_B - \mu_A$.
The basins $b \in \{A, B\}$ label the two coexisting phases (A-rich and B-rich).
Generalization to multi-species alloys follows by replacing $c$ with the simplex-valued composition $\bm{c}$ of~\cref{eq:cv} and $\Delta\mu$ with a vector of chemical-potential differences relative to a reference species.

For each experimental state point $m = (T_m, P_m)$, we run a single reference SGCMC plus WTMetaD simulation at parameters $\theta_0$ and chemical potential $\Delta\mu_{0,m}$ chosen close to coexistence so that both basins are sampled, and store $M_m$ decorrelated samples
\begin{equation}
\label{eq:difftre-samples}
    \mathcal{D}_m := \big\{ \bm{z}_{m,i},\, c_{m,i},\, V_{m,i},\, \kappa_{m,i} \big\}_{i=1}^{M_m},
\end{equation}
where $c_{m,i} := c(\bm{z}_{m,i})$, $V_{m,i} := V_\mathrm{bias}^{(m)}(c_{m,i}, t_i)$ is the metadynamics bias of~\cref{eq:Vbias} at the deposition time $t_i$, and $\kappa_{m,i}$ is the time-dependent reweighting offset of well-tempered reweighting~\citep{tiwary2015time} (an irrelevant additive constant if a frozen final bias is used).
The simulator is not differentiated; gradients flow only through reweighted observables on $\mathcal{D}_m$.

For a candidate $(\theta, \Delta\mu)$, the importance log-weight of sample $i$ relative to the reference is
\begin{multline}
\label{eq:difftre-logweight}
    \ell_{m,i}(\theta, \Delta\mu) := -\beta_m\, \big[ G_\theta(\bm{z}_{m,i};\, T_m, P_m) \\
    - G_{\theta_0}(\bm{z}_{m,i};\, T_m, P_m) - N\, (\Delta\mu - \Delta\mu_{0,m})\, c_{m,i} \big] \\
    + \beta_m\, (V_{m,i} - \kappa_{m,i}),
\end{multline}
with $\beta_m := 1 / k_\mathrm{B} T_m$.
The bracketed term is the semigrand-energy difference between candidate and reference parameters, and the $V_{m,i} - \kappa_{m,i}$ term unbiases the metadynamics bias accumulated during the reference simulation.
Normalizing by softmax,
\begin{equation}
\label{eq:difftre-weights}
    w_{m,i}(\theta, \Delta\mu) := \frac{\exp \ell_{m,i}(\theta, \Delta\mu)}{\sum_{j=1}^{M_m} \exp \ell_{m,j}(\theta, \Delta\mu)},
\end{equation}
yields a finite-sample estimator of any semigrand expectation under the candidate measure
\begin{equation}
\label{eq:difftre-expect}
    \widehat{\mathbb{E}}^{(m)}_{\theta, \Delta\mu}\!\left[ O \right] := \sum_{i=1}^{M_m} w_{m,i}(\theta, \Delta\mu)\, O(\bm{z}_{m,i}).
\end{equation}

The differentiable observable for binodal calibration is a basin-conditioned mean concentration, not the full free energy profile.
Define basin masks $M_{m,b}: [0, 1] \to \{0, 1\}$ on the concentration axis, with $M_{m,b}(c) = 1$ iff $c$ lies in the support of basin $b$ identified from the unbiased reference profile (typically the interval enclosing the corresponding minimum, with the boundary between basins placed near the intervening barrier).
The basin-local weights and predicted binodal concentration are
\begin{align}
\label{eq:difftre-basin-w}
    w_{m,i}^{(b)}(\theta, \Delta\mu) &:= \frac{M_{m,b}(c_{m,i})\, \exp \ell_{m,i}(\theta, \Delta\mu)}{\sum_j M_{m,b}(c_{m,j})\, \exp \ell_{m,j}(\theta, \Delta\mu)}, \\
\label{eq:difftre-cb}
    \widehat{c}_{m,b}(\theta, \Delta\mu) &:= \sum_i w_{m,i}^{(b)}(\theta, \Delta\mu)\, c_{m,i}.
\end{align}

At coexistence, $\Delta\mu$ is the value enforcing equal semigrand free energies of the two basins, so it is not a free parameter.
With unnormalized basin partition function
\begin{equation}
\label{eq:difftre-Zb}
    Z_{m,b}(\theta, \Delta\mu) := \sum_i M_{m,b}(c_{m,i})\, \exp \ell_{m,i}(\theta, \Delta\mu),
\end{equation}
coexistence reads $Z_{m,A}(\theta, \Delta\mu) = Z_{m,B}(\theta, \Delta\mu)$, or equivalently
\begin{equation}
\label{eq:difftre-coexist}
    H_m(\theta, \Delta\mu) := \ln Z_{m,A}(\theta, \Delta\mu) - \ln Z_{m,B}(\theta, \Delta\mu) = 0.
\end{equation}
For each $\theta$, \cref{eq:difftre-coexist} is solved for $\Delta \mu$ as a scalar root-finding problem using the stored samples~\cref{eq:difftre-samples} without additional simulation.
The condition~\cref{eq:difftre-coexist} is the ``common-tangent construction'' for the binodal: the two basins, evaluated at a common candidate $\Delta\mu$, share the same tangent slope, and equality of their semigrand free energies (equivalently $Z_{m,A} = Z_{m,B}$) makes that tangent ``touch'' both branches of the composition free energy.
Denoting the solution by $\Delta\mu_m^\star(\theta)$, the coexistence-conditioned binodal observable is
\begin{equation}
\label{eq:difftre-cstar}
    \widehat{c}_{m,b}^{\,\star}(\theta) := \widehat{c}_{m,b}\big( \theta,\, \Delta\mu_m^\star(\theta) \big).
\end{equation}
Implicit differentiation of~\cref{eq:difftre-coexist} yields the gradient of $\Delta\mu_m^\star$ with respect to $\theta$,
\begin{equation}
\label{eq:difftre-impl}
    \frac{\mathrm{d} \Delta\mu_m^\star}{\mathrm{d} \theta} = \frac{\widehat{\mathbb{E}}^{(m, A)}\!\left[ \nabla_\theta G_\theta \right] - \widehat{\mathbb{E}}^{(m, B)}\!\left[ \nabla_\theta G_\theta \right]}{N\, \big( \widehat{c}_{m,A}^{\,\star} - \widehat{c}_{m,B}^{\,\star} \big)},
\end{equation}
where $\widehat{\mathbb{E}}^{(m, b)}$ denotes the basin-local expectation with weights $w_{m,i}^{(b)}$ at $\Delta\mu = \Delta\mu_m^\star(\theta)$.

Given experimental binodal data $\{c_{m,b}^\mathrm{exp}\}$ across state points $m$ and basins $b \in \{A, B\}$, the model parameters are obtained by minimizing the weighted regression loss
\begin{multline}
\label{eq:difftre-loss}
    \mathcal{L}(\theta) := \lambda_\mathrm{bin} \sum_m \sum_{b \in \{A, B\}} \big( \widehat{c}_{m,b}^{\,\star}(\theta) - c_{m,b}^\mathrm{exp} \big)^2 \\
    + \lambda_\mathrm{drift} \sum_m \Delta_{\mathrm{drift},m}(\theta)^2 + \lambda_\mathrm{anchor}\, \lVert \theta - \theta_0 \rVert_2^2,
\end{multline}
with $c_{m,b}^\mathrm{exp}$ the experimental boundary concentrations.
The energy-drift term
\begin{equation}
\label{eq:difftre-drift}
    \Delta_{\mathrm{drift},m}(\theta) := \frac{1}{N M_m} \sum_{i=1}^{M_m} \big\lvert G_\theta(\bm{z}_{m,i}) - G_{\theta_0}(\bm{z}_{m,i}) \big\rvert
\end{equation}
measures the absolute per-atom change in the surrogate branch free energy between the reference parameters $\theta_0$ and the current $\theta$, averaged over the $M_m$ stored samples $\mathcal{D}_m$ of~\cref{eq:difftre-samples} at state point $m$.
This term limits the parameter update within each cycle and maintains overlap with the sampled distribution.
The final term is an $L_2$ penalty toward the same reference parameters $\theta_0$.
The parameters are optimized by Adam (learning rate $5\times10^{-6}$, gradient clipping at norm $5$) for $160$ gradient steps within each cycle, with weights $\lambda_\mathrm{bin} = 100$, $\lambda_\mathrm{drift} = 300$, and $\lambda_\mathrm{anchor} = 3\times10^{4}$ in the deployed \ce{Au}--\ce{Pt} calibration.
Within each calibration cycle, the stored samples are reused by reweighting without further simulation; between cycles, the reference SGCMC plus WTMetaD simulation is re-run at the updated parameters and coexistence estimate.
Between cycles, the metadynamics bias is initialized from the previous cycle.
Because successive parameter updates are small, fewer deposition steps are required to reconverge.

\section{Training Details}
\label{sec:expdetails}

\subsection{Multi-Fidelity Training}
\label{sec:training}

\Cref{table:lf-loss} lists the weights and per-response caps.

\begin{table}[!ht]
\centering
\caption{\textbf{Pre-training loss terms.} Each is a $L_1$ penalty on the residual values. The extensive quantities are normalized per atom; the intensive responses are not. The cap (in the units shown) bounds each per-sample contribution, and a global gradient-norm clip of $10$ applies to the summed loss.}
\label{table:lf-loss}
\begin{tblr}{colspec={llccc}, rowsep=0.5pt, colsep=4pt}
\toprule
Term & Units & Per atom & $\lambda$ & Cap \\
\midrule
$\mathcal{L}_G$ & eV~atom$^{-1}$ & yes & $1$ & --- \\
$\mathcal{L}_V$ & \AA$^3$~atom$^{-1}$ & yes & $0.25$ & --- \\
$\mathcal{L}_S$ & J~K$^{-1}$~(mol\,atoms)$^{-1}$ & yes & $0.005$ & $200$ \\
$\mathcal{L}_{C_P}$ & J~K$^{-1}$~(mol\,atoms)$^{-1}$ & yes & $0.003$ & $200$ \\
$\mathcal{L}_B$ & GPa & no & $0.003$ & $2000$ \\
$\mathcal{L}_\alpha$ & $10^{-5}$~K$^{-1}$ & no & $0.05$ & $200$ \\
\bottomrule
\end{tblr}
\end{table}

The trained thermodynamic adapter is small relative to the frozen encoder it augments.
The UMA representation (\texttt{UMA-S-1.1}) carries ${\sim}1.5\times10^{8}$ parameters, of which ${\sim}6\times10^{6}$ are active per structure through its mixture-of-experts routing.
The trainable thermodynamic adapter has ${\sim}2.1\times10^{5}$ parameters.
Of these, the analytic coefficient heads account for ${\sim}9\times10^{3}$.
Mid-training adds a ${\sim}4.4\times10^{4}$-parameter LoRA update~\citep{hu2022lora} (Methods), bringing the adapter to ${\sim}2.6\times10^{5}$ parameters in total, of which the LoRA update is about $17\%$.

\subsection{Analytic Head Constants and Numerical Evaluation}
\label{sec:head-constants}

The head uses two fixed global normalization scales, $T_\mathrm{max} = 4000$~K and $P_\mathrm{max} = 40$~GPa.
The reduced temperature $\tau = T/T_\mathrm{max}$ and the normalized pressure $P/P_\mathrm{max}$ each map their sampled range into $(0, 1]$.
Neither normalization scale is material specific, so the predicted surface is a function of structure alone.
All head quantities use a fixed unit system: per-atom energies in eV, temperatures in K, per-atom volumes in \AA$^3$, and pressures in GPa.
The reference coefficients $c_0$--$c_3$, $c_\mathrm{log}$, and $c_\mathrm{inv}$ are then per-atom energies, and $\theta_\mathrm{E}$ is a temperature.
The equation-of-state parameters are read as $V_{0,i}$ in \AA$^3$, $B_{0,i}$ in GPa, and $B'_{0,i}$ dimensionless.
The polynomial coefficients $a^{V_0}_k$, $a^{B_0}_k$, and $a^{B'_0}_k$ enter their softplus as dimensionless numbers.
In the Murnaghan volume integral the pressure enters only through the dimensionless ratio $B'_{0,i}P/B_{0,i}$, and the pressure-volume work, formed in \AA$^3$~GPa, is converted to eV with $1~\mathrm{eV}/\text{\AA}^{3} = 160.2~\mathrm{GPa}$ before it is added to the reference branch.
All remaining constants are fixed as follows: the normalized Einstein-temperature floor $\theta_{\min} = 10^{-4}$, the temperature floor $\tau_{\min} = 10^{-8}$ (both in units of $T_\mathrm{max}$), and the gate exponent $p = 4$ regularize the normalization and vibrational terms.
The equation-of-state floors $V_{0,\min} = 10^{-6}$~\AA$^3$, $B_{0,\min} = 10^{-3}$~GPa, and $B'_{0,\min} = 10^{-3}$, together with the Murnaghan argument floor $x_{\min} = 10^{-8}$, keep the equation of state well-defined.
The initialization bias $B_\mathrm{off} = 100$~GPa is added inside the $B_{0,i}$ softplus so the untrained bulk modulus starts near $100$~GPa.

\section{Holland--Powell Calibration and Benchmark Details}
\label{sec:hp-benchmark}

The experimental anchoring of ordered phases (Methods) uses the Holland--Powell \texttt{ds62} dataset~\citep{holland2011improved}: a predefined set of \ce{Si}--\ce{Al}--\ce{Mg}--\ce{Ca}--\ce{O} crystalline phases (\cref{table:hp-phases}) together with reactions among them (listed below).
For each phase, \texttt{ds62} provides the experimentally assessed single-phase properties at $298.15$~K and $1$~bar (the isothermal bulk modulus $B_0$, the thermal expansivity $\alpha$, the standard entropy $S(298~\mathrm{K})$, and the molar volume $V_0$).
Every reaction and polymorph contributes a Gibbs free energy difference target.
The training phases are the binary and ternary oxides above the divider in~\cref{table:hp-phases}.
The six quaternary oxides below the divider, together with the held-out reactions that form them, are excluded from fine-tuning and used only for evaluation.
Held-out accuracy therefore measures transfer from binary and ternary training phases to quaternary phases.
Phase abbreviations follow Holland and Powell~\citep{holland2011improved}.

\begin{table}[!ht]
\caption{\textbf{Holland--Powell benchmark phases.}     Binary and ternary phases used for training are listed above the divider; quaternary phases used only for evaluation are listed below.}
\label{table:hp-phases}
\begin{tblr}{colspec={lll}, rowsep=0.5pt, colsep=6pt}
\toprule
Abbrev. & Composition & Mineral \\
\midrule
q & \ce{SiO2} & quartz \\
trd & \ce{SiO2} & tridymite \\
crst & \ce{SiO2} & cristobalite \\
coe & \ce{SiO2} & coesite \\
stv & \ce{SiO2} & stishovite \\
per & \ce{MgO} & periclase \\
lime & \ce{CaO} & lime \\
cor & \ce{Al2O3} & corundum \\
fo & \ce{Mg2SiO4} & forsterite \\
mwd & \ce{Mg2SiO4} & wadsleyite \\
mrw & \ce{Mg2SiO4} & ringwoodite \\
en & \ce{Mg2Si2O6} & enstatite \\
mpv & \ce{MgSiO3} & Mg-perovskite \\
mak & \ce{MgSiO3} & akimotoite \\
cen & \ce{Mg2Si2O6} & clinoenstatite \\
pren & \ce{Mg2Si2O6} & protoenstatite \\
lrn & \ce{Ca2SiO4} & larnite \\
rnk & \ce{Ca3Si2O7} & rankinite \\
wo & \ce{CaSiO3} & wollastonite \\
cpv & \ce{CaSiO3} & Ca-perovskite \\
pswo & \ce{CaSiO3} & pseudowollastonite \\
wal & \ce{CaSiO3} & walstromite \\
cstn & \ce{CaSi2O5} & Ca-Si titanite \\
ky & \ce{Al2SiO5} & kyanite \\
and & \ce{Al2SiO5} & andalusite \\
sill & \ce{Al2SiO5} & sillimanite \\
sp & \ce{MgAl2O4} & spinel \\
\midrule
mont & \ce{CaMgSiO4} & monticellite \\
merw & \ce{Ca3MgSi2O8} & merwinite \\
ak & \ce{Ca2MgSi2O7} & akermanite \\
di & \ce{CaMgSi2O6} & diopside \\
py & \ce{Mg3Al2Si3O12} & pyrope \\
gr & \ce{Ca3Al2Si3O12} & grossular \\
\bottomrule
\end{tblr}
\end{table}

Reactions are written with the \texttt{ds62} phase abbreviations of~\cref{table:hp-phases}.

\medskip
\noindent
\begin{minipage}{\linewidth}
\emph{Training (15):}\\[3pt]
\begin{minipage}[t]{0.49\linewidth}
\begin{itemize}[nosep, leftmargin=1.5em]
\item 2\,per + q $\rightarrow$ fo
\item fo + q $\rightarrow$ en
\item lime + q $\rightarrow$ wo
\item cor + q $\rightarrow$ ky
\item cor + q $\rightarrow$ and
\item cor + q $\rightarrow$ sill
\item 2\,lime + q $\rightarrow$ lrn
\item 3\,lime + 2\,q $\rightarrow$ rnk
\end{itemize}
\end{minipage}\hfill
\begin{minipage}[t]{0.49\linewidth}
\begin{itemize}[nosep, leftmargin=1.5em]
\item lime + 2\,q $\rightarrow$ cstn
\item per + cor $\rightarrow$ sp
\item cen $\rightarrow$ en
\item pren $\rightarrow$ en
\item pswo $\rightarrow$ wo
\item and $\rightarrow$ ky
\item sill $\rightarrow$ ky
\end{itemize}
\end{minipage}
\end{minipage}

\medskip
\noindent
\begin{minipage}{\linewidth}
\emph{Polymorph (16):}\\[3pt]
\begin{minipage}[t]{0.49\linewidth}
\begin{itemize}[nosep, leftmargin=1.5em]
\item trd $\rightarrow$ q
\item crst $\rightarrow$ q
\item coe $\rightarrow$ q
\item stv $\rightarrow$ coe
\item mwd $\rightarrow$ fo
\item mrw $\rightarrow$ fo
\item 2\,mpv $\rightarrow$ en
\item 2\,mak $\rightarrow$ en
\end{itemize}
\end{minipage}\hfill
\begin{minipage}[t]{0.49\linewidth}
\begin{itemize}[nosep, leftmargin=1.5em]
\item cen $\rightarrow$ en
\item pren $\rightarrow$ en
\item cpv $\rightarrow$ wo
\item pswo $\rightarrow$ wo
\item wal $\rightarrow$ wo
\item and $\rightarrow$ ky
\item sill $\rightarrow$ ky
\item sill $\rightarrow$ and
\end{itemize}
\end{minipage}
\end{minipage}

\medskip
\noindent\emph{Held-out (6):}
\begin{itemize}[leftmargin=1.5em, topsep=3pt, partopsep=0pt, itemsep=0pt, parsep=0pt]
\item per + lime + q $\rightarrow$ mont
\item 3\,lime + per + 2\,q $\rightarrow$ merw
\item 2\,lime + per + 2\,q $\rightarrow$ ak
\item wo + per + q $\rightarrow$ di
\item 1.5\,en + cor $\rightarrow$ py
\item 3\,wo + cor $\rightarrow$ gr
\end{itemize}

The post-training objective is smooth-$L_1$ (Huber) throughout, apart from the $L_2$ weight regularizer $\mathcal{L}_\mathrm{reg}$.
The reaction, polymorph, and shape terms are normalized per atom.
All loss terms are evaluated on the training phases, and $\mathcal{L}_\mathrm{rxn}$ and $\mathcal{L}_\mathrm{poly}$ run over the $15$ training reactions and $16$ polymorph pairs above.
The single-phase anchors $\mathcal{L}_{B_0}$, $\mathcal{L}_{S_0}$, and $\mathcal{L}_{V_0}$ target the \texttt{ds62} bulk modulus, entropy, and molar volume tabulated at $298.15$~K and $1$~bar.
The remaining data terms target the \texttt{ds62} apparent Gibbs surface $G^\mathrm{ds62}(T,P)$.
A global gradient-norm clip of $5$ applies to the summed loss.
The per-term weights are collected in~\cref{table:hp-loss}.

\begin{table}[!ht]
\centering
\caption{\textbf{Post-training loss terms for ordered phases.} Per-term weights $\lambda$ of the experimental calibration objective (Methods).}
\label{table:hp-loss}
\begin{tblr}{colspec={llc}, rowsep=0.5pt, colsep=6pt}
\toprule
Term & Target & $\lambda$ \\
\midrule
$\mathcal{L}_\mathrm{rxn}$ & reaction $\Delta G$ & $1$ \\
$\mathcal{L}_\mathrm{poly}$ & polymorph $\Delta G$ & $0.5$ \\
$\mathcal{L}_{B_0}$ & $B_0$ & $0.003$ \\
$\mathcal{L}_{S_0}$ & $S(298~\mathrm{K})$ & $0.003$ \\
$\mathcal{L}_{V_0}$ & $V_0$ & $0.015$ \\
$\mathcal{L}_{V}$ & $V(T,P)$ & $0.05$ \\
$\mathcal{L}_{C_P}$ & $C_P(T)$ & $0.3$ \\
$\mathcal{L}_\mathrm{shape}$ & $G(T,P)-G(T_0,P_0)$ & $0.1$ \\
$\mathcal{L}_\mathrm{reg}$ & pre-calibration weights & $0.00001$ \\
\bottomrule
\end{tblr}
\end{table}

\bibliography{main}